\documentclass[fleqn,usenatbib,onecolumn]{mnras}

\usepackage{newtxtext,newtxmath}

\usepackage[T1]{fontenc}
\usepackage{ae,aecompl}

\newcommand{\rd}[1]{{#1}}

\usepackage{graphicx}	
\usepackage{amsmath}	
\usepackage{amssymb}	
\usepackage{times}

\usepackage{commath}
\usepackage{tikz,pgfplots}
\usepackage{upgreek}
\usepackage{natbib}

\usepackage{hyperref}
\usepackage{paralist}

\usetikzlibrary{pgfplots.groupplots}

\newcommand{\properbinom}[2]{\left({{#1}\atop#2}\right)}
\newcommand{\expe}{\mathrm{e}}
\newcommand{\Beta}{\mathcal{B}}

\title[Galaxy halo expansions]{Galaxy halo expansions: a new biorthogonal family of potential-density pairs}

\author[E. Lilley et al.]{
Edward J. Lilley,$^{1}$\thanks{E-mail: ejl44,jls,nwe@cam.ac.uk}
Jason L. Sanders,$^{1}$
N. Wyn Evans$^{1}$, Denis Erkal$^{2}$
\\
$^{1}$Institute of Astronomy, University of Cambridge, Madingley Road, Cambridge CB3 0HA, UK\\ $^{2}$ Department of Physics, Faculty of Engineering and Physical Sciences, University of Surrey, Guildford GU2 7XH, UK
}

\date{Accepted XXX. Received YYY; in original form ZZZ}

\pubyear{2017}

\begin{document}

\voffset=-0.6in

\label{firstpage}
\pagerange{\pageref{firstpage}--\pageref{lastpage}}
\maketitle

\begin{abstract}
  Efficient expansions of the gravitational field of (dark) haloes
  have two main uses in the modelling of galaxies:  first, they
  provide a compact representation of numerically-constructed (or
  real) cosmological haloes, incorporating the effects of triaxiality,
  lopsidedness or other distortion.  Secondly, they provide the basis
  functions for self-consistent field expansion algorithms used in the
  evolution of $N$-body systems.  We present a new family of
  biorthogonal potential-density pairs constructed using the Hankel
  transform of the Laguerre polynomials. The lowest-order density
  basis functions are double-power-law profiles cusped like $\rho \sim
  r^{-2 + 1/\alpha}$ at small radii with asymptotic density fall-off
  like $\rho \sim r^{-3 -1/(2\alpha)}$. Here, $\alpha$ is a parameter
  satisfying $\alpha \ge 1/2$. The family therefore spans the range of
  inner density cusps found in numerical simulations, but has much
  shallower -- and hence more realistic -- outer slopes than the
  corresponding members of the only previously-known family deduced by
  \cite{zhao1996} and exemplified by \citet{hernquist1992}.  When
  $\alpha =1$, the lowest-order density profile has an inner density
  cusp of $\rho \sim r^{-1}$ and an outer density slope of $\rho \sim
  r^{-3.5}$, similar to the famous \citet*{NFW1997} model. For this
  reason, we demonstrate that our new expansion provides a more
  accurate representation of flattened NFW haloes than the competing
  Hernquist-Ostriker expansion.  We utilize our new expansion by
  analysing a suite of numerically-constructed haloes and providing
  the distributions of the expansion coefficients.
\end{abstract}




\section{Introduction}

A galaxy is often modelled as a sum of simple analytical components,
such as a \citet{Hernquist1990} bulge or a \citet*[][NFW]{NFW1997} or
axisymmetric logarithmic halo ~\citep{Bi87, Ev93}.  Although crude,
such an approach produces manageable zeroth-order models from which
further insights can be garnered. However, in the era of powerful
numerical simulations of galaxy formation as well as accurate
positional and kinematic data in our own Milky Way, such broad-brush
models fail to capture the complexity of structures observed in both
simulations and data. A richer and more flexible way of representing
the gravitational field of galaxies (and their components) is to use
basis-function or halo-expansion techniques~\citep[see
  e.g.,][]{cluttonbrock1972, hernquist1992, zhao1996, lowing2011}. In
this framework, small higher-order deviations away from the
zeroth-order galaxy model are efficiently captured by terms in a
orthogonal series (analogous to a Fourier series), truncated at some
order $n_\mathrm{max}$. Separate sets of orthogonal basis functions
are used for the density and the potential, so the method is sometimes
known as the biorthogonal series expansion method. The two series are
related through the Poisson equation, enabling the same set of
coefficients to be used for both. This represents a significant
improvement in flexibility, allowing axisymmetric, triaxial, lopsided
or distorted density distributions to be built up from an underlying
simple model.

The original motivation for the introduction of biorthogonal basis
sets was the study of the stability of collisionless stellar
systems. When decomposed into basis functions, the normal modes of
stellar systems correspond to the eigenvectors of matrix equations
\citep[e.g.,][]{polyachenko1981,fridman84,weinberg1989,palmer1994,evansread1998}. Stability
studies of spherical galaxies often use spherical Bessel functions as
the radial basis functions and spherical harmonics as the angular
basis functions. These are natural choices as they are the
eigenfunctions of the Laplacian in spherical coordinates. However,
spherical Bessel functions form a discrete basis only if the radial
range is finite, so that the galaxy model has to be truncated or of
finite extent. For a semi-infinite radial range, \citet{saha1991} used
a transformation to map the radial range to a finite one, and then
used the brute force Gram-Schmidt method to build orthogonal basis
functions.

Subsequently, basis function techniques were used to provide
algorithms to evolve collisionless stellar systems, partly as a check
on the results of linear stability theory. For example,
\citet{Allen1990} used a biorthogonal spherical Bessel basis function
expansion to study the radial and circular orbit instability in
spherical galaxy models. As instabilities in stellar systems often
arise from nearly-resonant orbits, so accurate modelling of the
precession of individual orbits for several orbital periods is very
important. \citet{hernquist1992} clearly articulated the advantages of
basis function expansions for which the zeroth-order model actually
resembles a galaxy. They used the clever method pioneered by
\citet{cluttonbrock1972} to transform the (radial part of the)
Laplacian into an equation whose eigenfunctions are already
known. They tellingly remarked that "for reasons not immediately
obvious to us, Clutton-Brock's approach has been virtually ignored in
the literature." In the \citet{hernquist1992} basis function
expansion, the zeroth-order model is a \citet{Hernquist1990} density
profile, which resembles an elliptical galaxy or dark matter halo with
$\rho \sim 1/r$ at small radii and $\rho \sim 1/r^4$ at large radii.
  
Direct simulation of $N$-body gravitational systems requires $O(N^2)$
operations to calculate the forces on the particles. However, stellar
systems are typically collisionless, meaning two-body relaxation
effects between bodies are small and the particles chiefly move under
the influence of the smooth, mean gravitational field. It is this fact
that enables the use of efficient basis function expansion techniques
for calculating the force~\citep[see e.g.,][]{dehnen2014}. When forces
are calculated from the truncated series, we automatically avoid
spurious relaxation effects that may otherwise have been introduced
due to under-sampling of the number of particles in the system. The
coefficients are only calculated once (per time-step), and the series
of functions is evaluated once per particle to find the force, so such
methods are $O(N)$ in the number of particles.

More recently, basis function techniques have enjoyed a resurgence,
inspired by a very different motivation. \citet{lowing2011} pointed
out the value of describing the evolution of direct and
computationally expensive simulations of galaxy formation using basis
functions. Snapshots of the original simulation with redshift are
decomposed with basis functions. The entire simulation can then be
replayed many times at will. For example, new objects can be inserted
into the simulations and their behaviour studied as if they had been
present originally. This technique has already been exploited in
studies of the evolution of tidal streams in the
Galaxy~\citep{Ngan2015,Ngan2016}. However, the gamut of applications
is far broader, including the evolution of accreting subhaloes, the
dynamics of satellite galaxies and globular clusters in dark haloes and
the response of galactic disks to the changing dark halo potential.

Despite its significant advantages and diverse applications, an
important limitation of the basis function technique is the relative
paucity of available sets of `analytical' basis sets -- i.e. expressible
in closed form, and preferably obeying some form of recurrence
relation so that computation of the basis functions is efficient. Any
such basis sets are useful in their own right, but they are
particularly interesting if the zeroth-order potential-density pair is
close to that of galaxies, allowing for efficient representation using
only a few terms. In fact, both \citet{lowing2011} and
\citet{Ngan2015} decomposed numerical halo simulations into the
\citet{hernquist1992} basis function expansion, as the alternatives
are very limited.

The purpose of this paper is to broaden the application of the basis
function technique by introducing new sets of biorthogonal basis
functions and discussing their applications. We begin by reviewing the
presently available expansions and presenting a general approach to
the construction of biorthogonal expansions in
Section~\ref{Section::Theory} before presenting new specific cases in
Section~\ref{sec:construction}. Readers eager to apply the new basis
expansion can find the appropriate expressions at the beginning of
Section~\ref{sec::numerical}. We continue by expanding a series of
$N$-body cosmological haloes that are imitations of the Milky Way dark
halo using our new expansion method. We sum up and point out further
avenues of development, both on the theoretical and applicable side,
in Section~\ref{sec:conclusions}.

\section{Theory of biorthogonal expansions}\label{Section::Theory}

\subsection{History}

The story really begins with the two pioneering papers by Clutton-Brock. Seeking solutions to the cylindrical Poisson equation, \cite{cluttonbrock1972} introduced the Bessel functions as a set of biorthogonal functions with continuous eigenvalue on the interval $(0, \infty)$. From these functions, a biorthogonal set with discrete index $n$ was constructed by integrating a particular set of orthogonal functions $g_n(k)$ against this continuous eigenvalue (i.e. performing a Hankel transform). This procedure produced a basis set suitable for infinitesimally-thin disk galaxies. For Clutton-Brock's choice of $g_n(k)$ the lowest-order potential-density pair is that of the \citet{Kuzmin1956} or \citet{Toomre1963} disk, though he did not explicitly evaluate the integral and instead left the result as a recurrence relation. Subsequently, \cite{aoki1978} reproduced the results of \cite{cluttonbrock1972}, but they explicitly evaluated the Hankel transform to give results written in closed form.

For spherical systems, \cite{cluttonbrock1973} took a different approach, constructing a biorthogonal basis set for spherical galaxies directly out of the Gegenbauer polynomials together with a particular change of variable. The zeroth order density profile reproduced the \citet{Plummer1911} model, but this idea was then extended to cover the \citet{Hernquist1990} model by \cite{hernquist1992} before finally being generalised by \cite{zhao1996} to a family of basis sets for the hypervirial models \citep{Evans2005}. The extension of the \cite{cluttonbrock1972} approach has only received limited attention in the literature, but we shall demonstrate that it provides a powerful alternative route into constructing further basis sets. 

\cite{polyachenko1981} considered the Bessel functions over a finite interval $(0,a)$ where $a$ is the radial extent of the galaxy. This gives a biorthogonal basis set with discrete indices, although the expansion suffers from the issue that the Bessel functions by themselves do not resemble any particular well-known galactic profile. Unlike Clutton-Brock, they did not consider the case of a continuous eigenvalue on the semi-infinite interval $(0,\infty)$. They did however identify the additional degree of freedom, which allows for an adjustment in the asymptotic behaviour of the density basis functions and which we also exploit later.

More recently, \cite{rahmati2009} explicitly extended the derivation of \cite{cluttonbrock1972} and \cite{aoki1978} to the spherical case, making an analogous choice of $g_n(k)$, and hence deriving a basis set whose lowest order density is the perfect sphere \citep{dezeeuw1985}. Our work is a generalisation of this result, analogous to the way in which \cite{zhao1996} is a generalisation of \cite{cluttonbrock1973}. In fact, \cite{rahmati2009}'s basis set corresponds to the lowest member of our new family of biorthogonal pairs.

\subsection{Orthogonal Bessel solutions to the Poisson equation}

We seek solutions in series to the Poisson equation $\nabla^2\Phi =
4\pi G\rho$ (where we work with $G=1$ throughout). We represent the potential and density of a self-gravitating system
as a sum over basis functions
\begin{equation}
\Phi(\vec{r}) = \sum_{nlm} \: C_{nlm} \: \Phi_{nlm}(\vec{r}) \qquad\qquad\qquad
\rho(\vec{r}) = \sum_{nlm} \: C_{nlm} \: \rho_{nlm}(\vec{r}),
\end{equation} 
where the basis functions are labelled by the tuple $(n,l,m)$. The condition of biorthogonality is written
\begin{equation}\label{eq:biorth}
-\int \dif^{\:3} \vec{r} \: \Phi_{nlm} \: \rho^*_{n^\prime l^\prime m^\prime} = N_{nlm} \: \delta_{nlm}^{n^\prime l^\prime m^\prime}, \qquad\qquad\qquad
\nabla^2\Phi_{nlm} = 4\pi \rho_{nlm}.
\end{equation}
We factor out the spherical harmonics $Y_{lm}(\theta,\phi)$ as 
\begin{equation}
\Phi_{nlm}(\vec{r}) \equiv \Phi_{nl}(r) Y_{lm}(\theta,\phi),\qquad\qquad\qquad \rho_{nlm}(\vec{r})
\equiv \rho_{nl}(r) Y_{lm}(\theta,\phi),
\end{equation} 
leaving the equation satisfied by the radial basis functions as
\begin{equation}\label{eq:basis_function_equation}
\left(\nabla^2 - \frac{l(l+1)}{r^2}\right)\Phi_{nl} = 4\pi\rho_{nl}.
\end{equation}

On a finite domain with spherical symmetry, it is natural to choose as a basis the eigenfunctions of the Laplacian,
$f_{klm}(\vec{r}) = j_l(kr)\:Y_{lm}(\theta,\phi)$, where $j_l(kr) =
r^{-1/2}J_{l + 1/2}(kr)$ is a spherical Bessel function. The Bessel function of order $\mu$ is defined by the equation
\begin{equation}
-\frac{\dif}{\dif z} \left( z^{2\mu+1} \frac{\dif}{\dif z} \left( y_k(z) \right) \right) = k^2 z^{2\mu+1} y_k(z), \:\:\:\:\:\:\:\: y_k(z) \equiv z^{-\mu} J_\mu(kz),
\end{equation}
and the eigenfunction $f_{klm}$ obeys the equation $\nabla^2f_{klm} = k^2\:f_{klm}$. However, on an infinite domain we cannot use these eigenfunctions directly for a series expansion as the radial eigenvalue $k$ becomes continuous: every element of $(0,\infty)$ is a valid eigenvalue $k$. We therefore convert the basis to a discrete one; first choosing a set of functions $g_n(k)$ that are orthogonal on the interval $(0,\infty)$ with respect to the measure $k \dif k$ and then integrating against $f_{klm}$ (essentially performing a Hankel transform). This produces a \emph{biorthogonal} basis set, that is suitable for expanding both the potential $\Phi$ and the density $\rho$. 

Given that we are using different sets of functions to represent the potential and the density, we can exploit an additional freedom first noticed by \citet{polyachenko1981} and introduce the parameter
$\alpha$ in the $r$-dependence of the Bessel functions. Accordingly, we define the potential and density basis functions as
\begin{equation}
\Phi_{nl}(r) \equiv \frac{-A_{nlm}}{\sqrt{r}} \int_0^\infty \dif k \: g_n(k) \: J_{\mu}\left(kr^{1/(2\alpha)}\right) \: 
, \qquad\qquad\qquad
\rho_{nl}(r) \equiv \frac{A_{nlm}}{16\pi\alpha^2} r^{1/\alpha - 5/2} \int_0^\infty \dif k \: k^2 \: g_n(k) \: J_{\mu}\left(kr^{1/(2\alpha)}\right) \: ,
\label{eq:generic_basis}
\end{equation}
where $\mu \equiv \alpha \: (1 + 2l)$, and
$A_{nlm}$ are constants that will be adjusted to give the desired overall normalisation. These functions are biorthogonal in the sense of \eqref{eq:biorth} if the auxiliary functions $g_n(k)$ are suitably chosen. Specifically, they must satisfy an orthogonality condition which can be derived thus,
\begin{equation}\label{eq:orthog}
\begin{split}
-\int_0^\infty \dif r \: r^2 \: \Phi_{nl} \: \rho_{n^\prime l} 
& = \frac{A_{nlm}\:A_{n^\prime lm}}{8\pi\alpha} \: \int_0^\infty \dif k \: g_n(k) \int_0^\infty \dif q \: q^2 g_{n^\prime}(q) \int_0^\infty \dif z \: z \: J_{\mu}(k z) J_{\mu}(q z) \qquad\qquad\qquad \left(z \equiv r^{1/(2\alpha)}\right) \\
& = \frac{A_{nlm}\:A_{n^\prime lm}}{8\pi\alpha} \: \int_0^\infty \dif k \: k \: g_n(k) \: g_{n^\prime}(k) =  \frac{A_{nlm}^2}{8\pi\alpha} \: I_n \: \delta_{n n^\prime}.
\end{split}
\end{equation}
Here $I_n$ is the normalisation constant for the functions $g_n(k)$ are orthogonal with respect to the weight function $k$. Let us also write $J_{lm}$ as the normalisation constant for the spherical harmonics $Y_{lm}$. The overall normalisation constant is
therefore 
\begin{equation}
N_{nlm} = \frac{I_n \: J_{lm} \: A_{nlm}^2}{8\pi\alpha},
\end{equation}
so that the functions are biorthonormal (i.e. $N_{nlm}=1$) if we set 
\begin{equation}
A_{nlm} =
\sqrt{\frac{8\pi\alpha}{I_n\:J_{lm}}},
\end{equation}
in eq.~\eqref{eq:generic_basis}. The coefficients $C_{nlm}$ can be calculated by the following integrals over $\rho$ and $\Phi$. When the density is formed from a cloud of point particles as in a numerical simulation, the integral reduces to a sum, which (given this choice of normalisation) is written
\begin{equation}\label{eq:particle_sum}
\rho(\vec{r}) = \sum_i\:m_i \delta^3(\vec{r} - \vec{r_i}), \qquad\qquad\qquad C_{nlm} = -\int \dif^{\:3}\vec{r} \: \Phi_{nlm}(\vec{r})\:\rho(\vec{r}) = \sum_i m_i\:\Phi_{nlm}(\vec{r_i}).
\end{equation}
One of the advantages of biorthogonal basis sets is that the coefficients in both the density and the potential are the same and can be evaluated very easily from an $N$-body realisation. Note that the use of a discrete sum instead of a continuous integration implies that there is always some numerical (Poisson) noise introduced by discreteness effects.

\subsection{Physical boundary conditions}\label{sec:asymp}

The approach presented above yields general sets of biorthogonal functions, but there are further considerations when wishing to construct basis sets for physical systems. Consider the following argument about the required asymptotic behaviour of the radial basis functions. For a given radial order $n$ and angular order $l$, we define the mass enclosed by a shell with inner radius $a$ and outer radius $b$ as
\begin{equation}\label{eq:encmass}
M_{nl}(a, b) = 4\pi \int_a^b \dif r \: r^2 \: \rho_{nl} = \int_a^b \dif r \: \left[\frac{\dif}{\dif r}\left(r^2\frac{\dif}{\dif r}\Phi_{nl}\right) - l(l+1)\Phi_{nl}\right].
\end{equation}
In order for the density and potential to describe a physical system, we impose the constraint that the mass enclosed by a thin shell goes to zero at the origin and at infinity, and that the potential remains finite everywhere. That is, we require that
\begin{equation}\label{eq:limits}
\lim_{a\to 0} M_{nl}(0,a) = 0, \qquad\qquad\qquad \lim_{a\to\infty} M_{nl}(a,\infty) = 0, \qquad\qquad\qquad \Phi_{nl}(r) < \infty.
\end{equation}
These translate into equations that must be satisfied by the derivative of $M_{nl}$, which is proportional to $\rho_{nl}$, and so examining \eqref{eq:encmass} we see that the asymptotic solutions for the potential must satisfy : $\Phi_{nl} \sim r^l$ as $r\to 0$; and $\Phi_{nl} \sim r^{-1-l}$ as $r \to \infty$.

\section{Construction of the new basis set}\label{sec:construction}

\subsection{Zeroth-Order Models}

With our general framework in place, we now turn to the job of producing expansions whose zeroth-order model resembles realistic galactic components. This is a lengthy business, so we begin by describing some of the zeroth-order models in our sequence of basis function expansions. Of course, the zeroth-order models are all spherically symmetric, but the higher-order terms in the basis function expansion describe the effects of flattening or lopsidedness or distortion. Hence, the method can build very general density distributions around the zeroth-order model.

The models are labelled by a parameter $\alpha$; however, those with $0< \alpha< 1/2$ have vanishing density at the origin, and so we consider them no further. Models of interest in galactic dynamics have $\alpha \ge 1/2$. The behaviour of the density at small and large radii is 
\begin{equation}
\lim_{r\to 0} \rho_{000} (r)  \sim r^{-2 + 1/\alpha}, 
\qquad\qquad \lim_{r\to\infty} \rho_{000} (r) \ \sim r^{-3 -1/(2\alpha)},
\end{equation}
so that the $\alpha = 1/2$ model is cored, whilst the remainder are cusped.  The density of all the models falls off with a logarithmic slope between $-3$ and $-4$. We proceed to list some of the zeroth-order models that are obtained when $\alpha$ is integer or half-integer, as in these cases the potential reduces to elementary functions; in general real values of $\alpha$ are permitted, but the potential must be evaluated using the incomplete beta function (see Eq.~\eqref{eq:lowest_orders}).

When $\alpha = 1/2$, the zeroth-order model is the {\it perfect sphere} of \cite{dezeeuw1985}, which has density and potential
\begin{equation}
\rho_{000} = \frac{1}{\sqrt{2} \: \pi^{3/2}} \: \frac{1}{\left(1 + r^2\right)^{2}},
\qquad\qquad\qquad\qquad\qquad\qquad
\Phi_{000} = -\sqrt{2\over \pi} \frac{\arctan{r}}{r}. 
\end{equation}
This is the spherical limit of the perfect ellipsoid, which provides triaxial densities that are close to the luminosity profiles of elliptical galaxies. There is a large literature on these models as the potential is separable or St\"ackel, and so the orbits and action-angles can be found as quadratures~\citep[e.g.,][]{Bi87,Be14}. Notice that the density has a harmonic core at the centre but falls like $\rho \sim r^{-4}$ at large radii.

When $\alpha = 1$, we obtain the {\it super-NFW model}
\citep[e.g.,][]{Ev06,An13,Lilley2017b}. This has a central density cusp with $\rho
\propto r^{-1}$ similar to the \citet{Hernquist1990} or NFW
models. Asymptotically, the density falls like $\rho \propto r^{-3.5}$
at large radii, which is somewhat faster than the NFW model, but has
the distinct advantage of finite total mass.  There is evidence 
to suggest that the NFW may be too shallow at large radii based on the
recent work on the splashback radius by~\citet{Di17}. The potential-density
pair of the super-NFW model is
\begin{equation}
\rho_{000} = \frac{3}{16\pi} \: \frac{1}{r\left(1 + r\right)^{5/2}},
\qquad\qquad\qquad\qquad\qquad\qquad 
\Phi_{000} = \frac{-1}{1 + r + \sqrt{1 + r}}.
\end{equation}
This is ideal for modelling the profiles of dark haloes found in
cosmological simulations. This potential-density pair may also be used
as a simple model of a galaxy or star cluster. \citet{Lilley2017b}
show that isotropic and anisotropic distribution functions and
solutions of the Jeans equations can be obtained, whilst the projected
density is close to de Vaucouleurs and Sersic profiles. \rd{There are
  of course other finite mass models which have such desirable
  properties. A number of models with $\rho \propto r^{-1}$ at the
  centre and with an asymptotic fall off with logarithmic gradient
  between -3 and -4 are already available in the literature
  \citep[e.g.,][]{An06,An13}. In general, these models are not
  the zeroth-order terms of any biorthorgonal expansion, so their
  axisymmetric and triaxial siblings are much more cumbersome to
  construct.}

More strongly cusped models can also be obtained as zeroth-order models. When $\alpha = 3/2$, we have
\begin{equation}
\rho_{000} = \frac{2\sqrt{2}}{9 \pi^{3/2}} \: \frac{1}{r^{4/3}\left(1 + r^{2/3}\right)^{3}},
\qquad\qquad\qquad\qquad\quad
\Phi_{000} = \sqrt{2\over \pi}\Bigl[ {1\over r^{2/3}(1 + r^{2/3})} - {{\arctan} (r^{1/3})\over r}\Bigr].
\end{equation}
The central density cusp is now $\rho \sim r^{-4/3}$, similar to the inner regions of the massive cosmological clusters studied by \citet{Diemand2005}. When $\alpha = 2$, we have
\begin{equation}
\rho_{000} = \frac{15}{64 \pi} \: \frac{1}{r^{3/2}\left(1 + r^{1/2}\right)^{7/2}},
\qquad\qquad\qquad\qquad\qquad
\Phi_{000} = -{1 + 2\sqrt{1+\sqrt{r}} \over \left(1 + \sqrt{1+\sqrt{r}}\right)^2 \left(1+\sqrt{r}\right)^{3/2}}.
\end{equation}
This has an inner density profile $\rho \sim r^{-3/2}$, similar to the very steepest cusps found in cosmological simulations \citep{Moore98}.

Explicitly, the lowest-order basis functions are
\begin{equation}\label{eq:lowest_orders}
\rho_{000} = {2^{\alpha+1}\Gamma(3/2+\alpha)\over\sqrt{\pi}}
{1 \over r^{2 - 1/\alpha}(1+r^{1/\alpha})^{\alpha+ 3/2}}\qquad\qquad\qquad
\Phi_{000} = -{2^{\alpha-1} \Gamma(\alpha + 3/2) \over \sqrt{\pi}} {\Beta_\chi(\alpha,1/2) \over r}, \qquad\qquad  \chi \equiv \frac{r^{1/\alpha}}{1 + r^{1/\alpha}},
\end{equation}
where $\Beta_x(a,b)$ is the incomplete beta function. The full suite of models has a range of inner density profiles suitable for representing dark haloes with cores and weak or strong cusps, and $\alpha$ can be tuned to match the behaviour of a given halo (see Sec.~\ref{Section::Numhaloes}).

\subsection{Choice of Auxiliary Function}

Now that we have given the spherically-symmetric lowest-order models, we show how construct the associated higher-order basis functions. We introduce a two parameter family of auxiliary functions $g_n(k)$. We derive an expression for the potential basis functions but find that the requirement of physicality reduces this set to a single parameter family. This reduction of complexity allows us to express the potential and density basis functions in a more succinct fashion in Sections~\ref{sec::potential} and ~\ref{sec::density}.

Inspired by \cite{rahmati2009}\footnote{Who were in turn motivated by the analogous disc case considered in \cite{cluttonbrock1972}.}, and initially following their calculations, we now choose
\begin{equation}\label{eq:gnk_def}
g_n(k) \equiv L^{(\eta)}_n(2k) \: k^{(\eta-1)/2} \: \exp(-k),
\end{equation}
where $L^{(\eta)}_n(x)$ are the associated Laguerre polynomials \citep[\S 18.3]{dlmf}, and $\eta$ is (for now) a free parameter. The Laguerre polynomials are orthogonal on $(0,\infty)$ with respect to a weight function $\omega(x) \equiv x^\eta \: \exp (-x)$,
\begin{equation}
\int_0^\infty \dif x \: L^{(\eta)}_n(x) \: L^{(\eta)}_{n^\prime}(x) \omega(x) = \delta_{n n^\prime} {\Gamma(n+\eta+1)\over \Gamma(n+1)}\:.
\end{equation}
Our $g_n(k)$ consist of a Laguerre polynomial multiplied by a factor of $\sqrt{\omega(2k)/k}$, hence ensuring the orthogonality on $(0,\infty)$ with respect to $k \dif k$. We can obtain an expression for $\Phi_{nl}$ using the integral Eq.~6.621(1) of \citet{gradshteyn4},
\begin{equation}\label{eq:hankel_alg}
\int_0^\infty \dif u \: \exp(-us) u^\nu J_\mu(u) = \Gamma(\nu + \mu + 1) \: (1 + s^2)^{-(1 + \nu)/2} \: P^{(-\mu)}_\nu\left(\frac{s}{\sqrt{1 + s^2}}\right).
\end{equation}
Setting $u = kr^{1/(2\alpha)}$ and $s = r^{-1/(2\alpha)}$, and using the explicit polynomial representation of the Laguerre polynomials
\begin{equation}\label{eq:laguerre}
L^{(\eta)}_n(x) = \sum_{j=0}^n (-1)^j \frac{1}{j!} \properbinom{n+\eta}{n-j} x^j,
\end{equation}
we find
\begin{equation}\label{eq:phi_alf}
\Phi_{nl} = \sum_{j=0}^n A_{n j\mu\eta} \left[ r \left(1 + r^{1/\alpha}\right)^{(\eta+1)/2 + j}\right]^{-1/2} \: P_{\frac{\eta-1}{2} + j}^{\left(-\mu\right)}\left(\frac{1}{\sqrt{1 + r^{1/\alpha}}}\right), \qquad\qquad\qquad
A_{n j\mu\eta} \equiv \frac{(-2)^j}{j!} \properbinom{n+\eta}{n-j} \Gamma\left(j + \frac{\eta+1}{2} + \mu\right).
\end{equation}
We use the results of section \ref{sec:asymp} on asymptotic limits to fix the parameter $\eta$. The asymptotic behaviour of the associated Legendre function \citep[Eq.~14.8(i)]{dlmf} as $z \to 1$ from above is $P^{(\mu)}_\nu(z) \sim (1 - z)^{-\mu/2}$. Therefore, as $r \to 0$, every term in $\Phi_{nl}$ goes as
\begin{equation}
\Phi_{nl} \sim r^{-1/2} \: \left(1 - \frac{1}{\sqrt{1 + r^{1/\alpha}}}\right)^{\mu/2} \propto r^{\mu/(2\alpha) - 1/2} = r^l,
\end{equation}
so the first limit implied by \eqref{eq:limits} is already satisfied. On the other hand, as $r \to \infty$, $1/\sqrt{1 + r^{1/\alpha}} \sim r^{-1/(2\alpha)}$, so the associated Legendre function goes to a constant \citep[Eq.~14.5.1]{dlmf} and we are left with the prefactor
\begin{equation}
\Phi_{nl} \sim \left[ r \left(r^{1/\alpha}\right)^{(\eta+1)/2 + j}\right]^{-1/2} \rightarrow r^{-1/2 - (\eta + 1)/(4\alpha)},
\end{equation}
where we have used the fact that the $j = 0$ term dominates the sum as $r \to \infty$. By \eqref{eq:limits}, we require this limit to be $r^{-l-1}$, which we can achieve by setting $\eta = 2\mu - 1$, so the $g_n$-normalisation constant is 
\begin{equation}
I_n = \frac{\Gamma(n+2\mu)}{2^{2\mu}\:n!}.
\end{equation}
Now that $\eta$ is no longer a free parameter, we can obtain the basis functions in their most convenient form.

\subsection{Recurrence Relation for The Potential Basis Functions}\label{sec::potential}

The sum \eqref{eq:phi_alf} is unsatisfactory for several reasons:
\begin{inparaenum}
\item it is numerically unstable for high $n$,
\item$n$ and $j$ are coupled in such a way that calculating $n$ basis functions requires $O(n^2)$ operations , and 
\item associated Legendre functions of non-integer or negative degree or order are rarely implemented numerically. 
\end{inparaenum}

Therefore, departing from \cite{rahmati2009}, we seek a superior expression, which can be obtained using the recurrence relation for the Laguerre polynomials,
\begin{equation}\label{eq:laguerre_recurrence}
n \: L_n^{(\alpha)}(x) = (n + \alpha) \: L_{n-1}^{(\alpha)}(x) - x \: L_{n-1}^{(\alpha+1)}(x).
\end{equation}
Using Eq.~(\ref{eq:gnk_def}), we can immediately write down
\begin{equation}
n \: g_n(k) = (n + 2\mu - 1) \: g_{n-1}(k) - 2 \exp (-k) \: k^\mu \: L_{n-1}^{(2\mu)}\left(2k\right).
\end{equation}
So, we obtain the following recurrence relation for the basis functions
\begin{equation}
n \: \Phi_{nl} = (n + 2\mu - 1) \: \Phi_{n-1,l} +
{2A_{nlm}\over \sqrt{r}} \int_0^\infty k^\mu \: L_{n-1}^{(2\mu)}(2k) \: \exp(-k) \: J_\mu(kz) \: \dif k,
\qquad\qquad\qquad \left(z \equiv r^{1/(2\alpha)}\right).
\end{equation}
To evaluate the latter integral is some work. First, we note that a generating function for the Laguerre polynomials is
\begin{equation}\label{eq:gen_lag}
\sum_{n=0}^\infty \: t^n \: L_n^{(\lambda)}(k) = \frac{\exp \left(-tk/(1-t) \right)}{\left(1-t\right)^{\lambda+1}},
\end{equation}
so that, using the following Hankel transform \citep[Eq.~6.623(1)]{gradshteyn4}
\begin{equation}
\int_0^\infty \: x^\nu \: \expe^{-ax} \: J_\nu(xy) \: \dif x = \frac{2^\nu \: \Gamma(\nu + 1/2)}{\sqrt{\pi}} \: \frac{y^\nu}{\left(a^2 + y^2\right)^{\nu + 1/2}},
\end{equation}
as well as the generating function for the Gegenbauer polynomials
$C_n^{(\lambda)}(\xi)$
\begin{equation}\label{eq:gen_gegen}
\sum_{n=0}^\infty \: t^n \: C_n^{(\lambda)}(\xi) = \frac{1}{\left(1 - 2\xi t + t^2\right)^\lambda} = \frac{\left(1 + z^2\right)^\lambda}{\left[(1+t)^2 + (1-t)^2z^2\right]^\lambda}, \qquad\qquad\qquad \left(\xi \equiv \frac{z^2 - 1}{z^2 + 1}
\equiv \frac{r^{1/\alpha} - 1}{r^{1/\alpha} + 1}\right),
\end{equation}
we can find the Hankel transform of the remaining term in the recurrence relation,
\begin{align}\label{eq:phi_generating_function}
\sum_{n=0}^\infty \: t^n \: \int_0^\infty k^\mu \: L_n^{(2\mu)}(2k) \: \exp (-k) \: J_\mu(kz) \: \dif k & = \int_0^\infty k^\mu \frac{\exp(-k(1+t)/(1-t))}{\left(1-t\right)^{2\mu + 1}}J_\mu(kz)\:\dif k 
 = \frac{2^\mu \: \Gamma(\mu + 1/2)}{\sqrt{\pi}} \: \frac{z^\mu}{\left[(1+t)^2 + (1-t)^2 z^2\right]^{\mu + 1/2}} \nonumber \\
& = \sum_{n=0}^\infty t^n \: \frac{2^\mu \: \Gamma(\mu + 1/2)}{\sqrt{\pi}} \: \frac{z^\mu}{\left(1+z^2\right)^{\mu + 1/2}} \: C_n^{(\mu + 1/2)}(\xi).
\end{align}
So the recurrence relation becomes
\begin{equation}\label{eq:recurrence}
n \: \Phi_{nl} = (n + 2\mu - 1) \: \Phi_{n-1,l} + A_{nlm} \frac{2^{\mu+1} \: \Gamma(\mu+1/2)}{\sqrt{\pi}} \: \frac{r^l \: C_{n-1}^{(\mu+1/2)}\left(\xi\right)}{\left(1 + r^{1/\alpha}\right)^{\mu+1/2}}.
\end{equation}
This means that, in theory, the $(n+1)$-th basis function can be trivially calculated from the $n$-th by simply adding on a single extra term. Evaluating $n$ basis functions therefore requires $O(n)$ steps (although see Section~\ref{sec:implementation}). A similar approach is taken in \cite{aoki1978} in the disk setting. 

So far we've only stated without proof the $n=0,l=0$ case. We now obtain an explicit expression for all the potential basis functions, from which the zeroth order ($n=0$, $l\geq 0$) can be extracted. We begin by writing our auxiliary function as
\begin{equation}\label{eq:gnk_separation}
g_n(k) = k^{\mu - 1} \: \exp(-k) \: L_n^{(2\mu - 1)}(2k) = k^{\mu - 1} \: \exp (-k) \left( L_n^{(2\mu - 1)}(0) - 2k\sum_{j=0}^{n-1} \frac{\properbinom{n+2\mu - 1}{n-1-j}}{(n-j)\properbinom{n}{j}} L_j^{(2\mu)}(2k)\right),
\end{equation} 
where again $\mu = \alpha \: (1+2l)$. Aside from the constant term, the Hankel transforms can be done using Eq.~(\ref{eq:phi_generating_function}). To evaluate the Hankel transform of the constant term, we use Eq.~\eqref{eq:hankel_alg} to find (denoting by $\Beta_x(a,b)$ the incomplete beta function)
\begin{equation}\label{eq:potential_zeroth_order}
\int_0^\infty \: k^{\mu-1} \: \exp(-k) \: J_\mu(kz) \: \dif k = \frac{\Gamma(2\mu)}{\left(1+z^2\right)^{\mu/2}} \: P^{(-\mu)}_{\mu-1}\left(\frac{1}{\sqrt{1+z^2}}\right) = \frac{\Gamma(2\mu)}{\Gamma(\mu)\:2^\mu\:z^\mu} \: \Beta_\chi(\mu,1/2), \qquad\qquad\qquad \left(\chi \equiv \frac{r^{1/\alpha}}{1 + r^{1/\alpha}}\right).
\end{equation}
%
%
%
%
%
%
%
%
%
So, reassembling the terms, we obtain an expression for the potential as
\begin{equation}\label{eq:phi_nl_def}
\Phi_{nl}(r) = -A_{nlm} \: \frac{2^\mu \: \Gamma(\mu + 1/2) \: \Gamma(n + 2\mu)}{\sqrt{\pi} \: n!} \left[\frac{\Beta_\chi(\mu,1/2)}{2\:\Gamma(2\mu)\:r^{1+l}} - \frac{2r^l}{\left(1 + r^{1/\alpha}\right)^{\mu+1/2}}\sum_{j=0}^{n-1}\frac{j! \: C_j^{(\mu+1/2)}(\xi)}{\Gamma(j + 2\mu + 1)}\right],\qquad\qquad\qquad \left( \chi \equiv \frac{1+\xi}{2}\right),
\end{equation}
with $\mu \equiv \alpha\:(1+2l)$. The only special functions required to evaluate the potential are the incomplete beta function and the Gegenbauer polynomials, which are standard library functions in many numerical software packages (both are included in the GNU Scientific Library, for example). Using the results above, we see that it is easy to write down and compute a basis set for all real values of $\alpha$.

\begin{figure}
$$\includegraphics[width=0.8\textwidth]{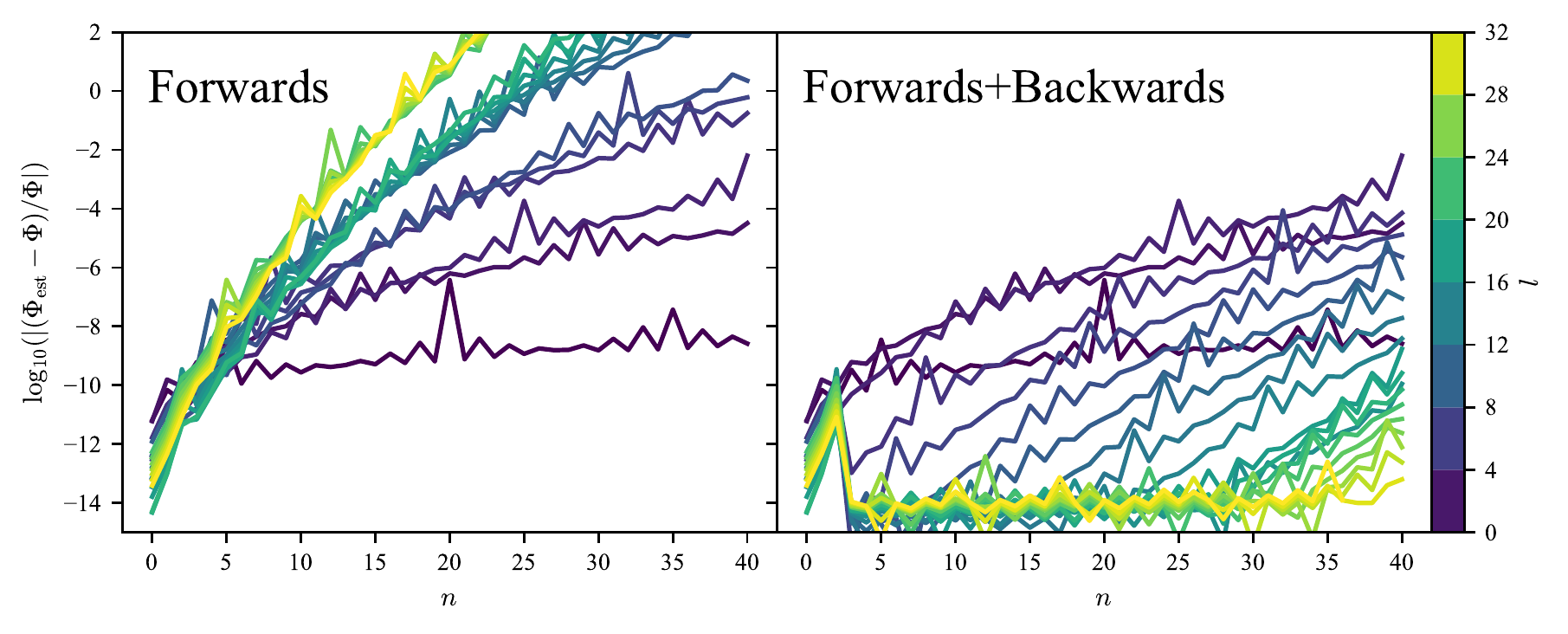}$$
\caption{Relative error in potential basis function calculation using
  two methods: on the left the result of using a forward recursion
  scheme and on the right using a combination of forward and backward
  recursion. \rd{The relative error is the difference between the
    potential computed by the basis expansion and an arbitrary
    precision calculation from {\it Mathematica} computed at $r/r_s=1.3$ using $\alpha=1.2$. Note that the floor
    on the error in the right panel is set by the precision of our
    {\it Mathematica} calculation and can be lowered if so desired. We only perform the `forwards+backwards' procedure for $l>4$.}}
\label{fig::numerical_accuracy}
\end{figure}

\begin{figure}
  \centering
  \includegraphics{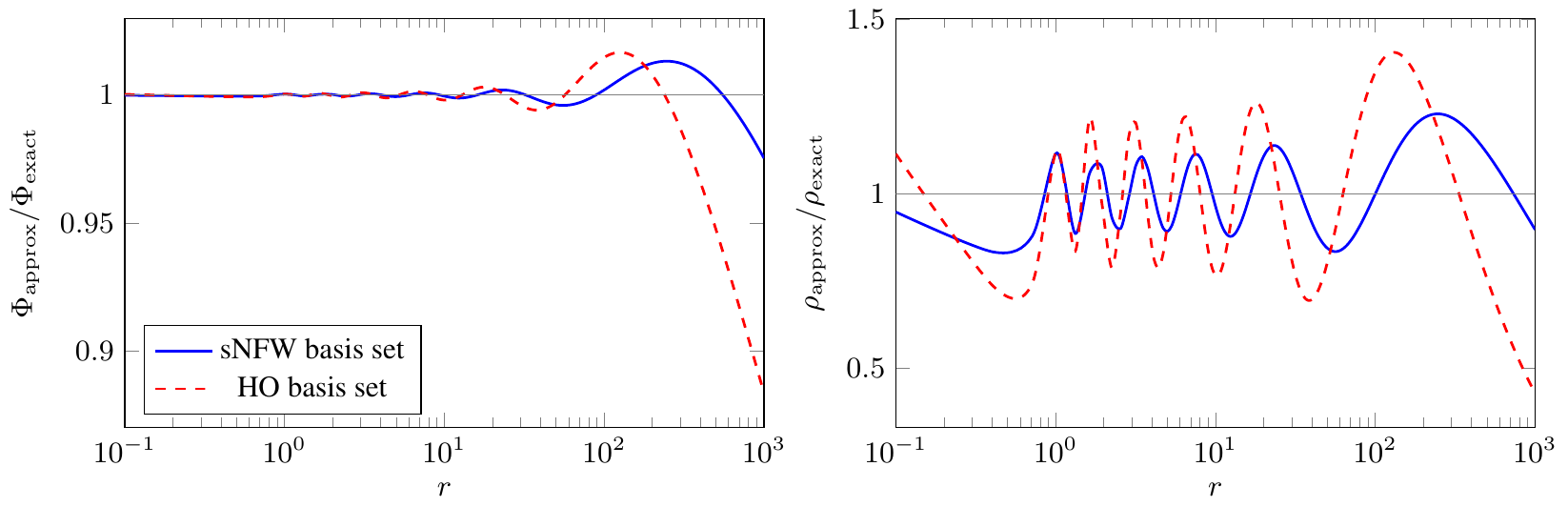}
  \caption{Reconstruction of a spherical NFW potential (left) and
    density (right) with the new \rd{super-NFW} basis set (blue) and
    the Hernquist-Ostriker basis set (red). The distance is given in
    units of the NFW scalelength. Both expansions use radial terms up
    to $n = 20$ and no angular terms ($l=0$). Both expansions
    oscillate around the true value, which is represented by the
    horizontal grey line.}\label{fig:spherical}
\end{figure}
\begin{figure}
  \centering
  \includegraphics{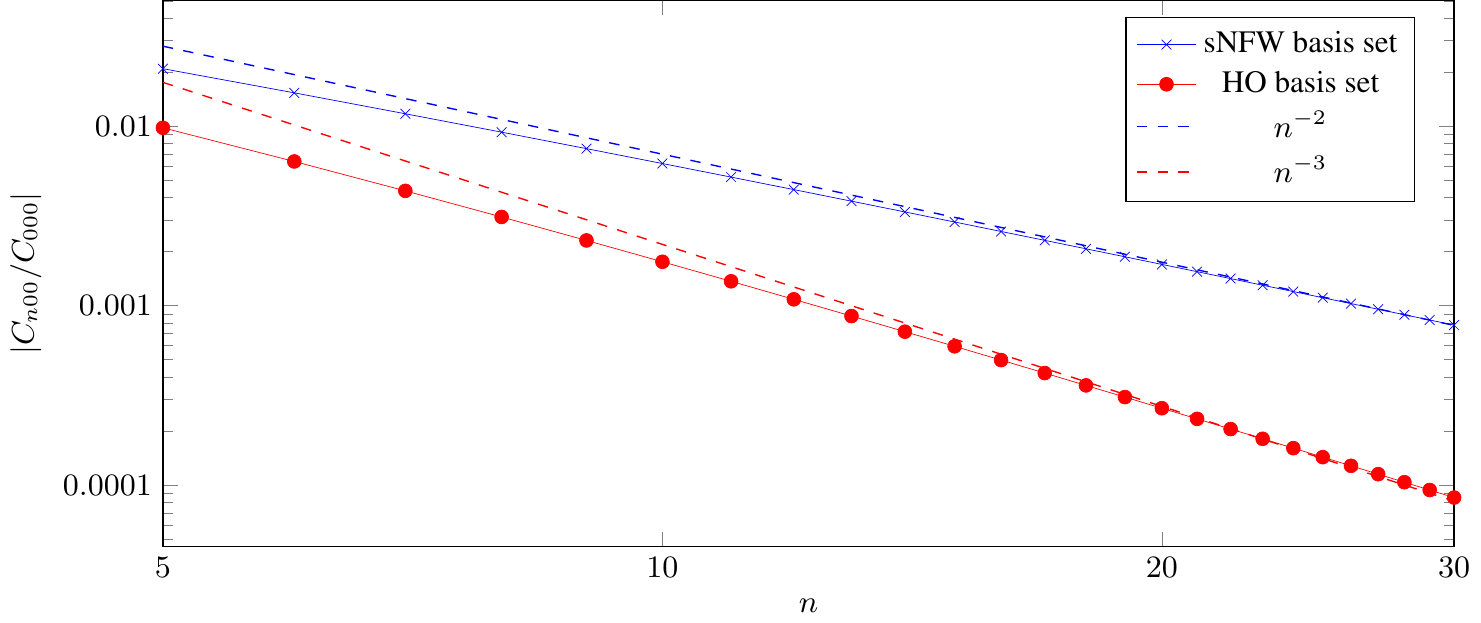}
  \caption{The run of the radial expansion coefficients with $n$ for a spherical NFW model using the new basis set (blue) and the Hernquist-Ostriker basis set (red). Also plotted as dashed lines are the $n^{-2}$ and $n^{-3}$ curves, suggesting that the coefficients fall off asymptotically like $n^{-2}$ in our case and like $n^{-3}$ for the Hernquist-Ostriker case.}\label{fig:coeffs}
\end{figure}

\subsection{The Density Basis Functions}\label{sec::density}

A similar calculation can be performed to find expressions for the density basis functions, though it is simpler as the generating function can be applied immediately with no further manipulation of $g_n(k)$. First, we note the following Hankel transform \citep[Eq.~6.623(2)]{gradshteyn4}
\begin{equation*}
\int_0^\infty x^{\nu+1} \: \exp (-ax ) \: J_\nu(xy) \: \dif x = \frac{2^{\nu+1} \: \Gamma(\nu + 3/2)}{\sqrt{\pi}} \: \frac{a \: y^\nu}{\left(a^2 + y^2\right)^{\nu + 3/2}}.
\end{equation*}
Then, using \eqref{eq:gen_lag} and \eqref{eq:gen_gegen} as above, we have
\begin{align}
\sum_{n=0}^\infty t^n \: \rho_{nl} & \propto r^{1/\alpha - 5/2} \: \sum_{n=0}^\infty \: t^n\int_0^\infty k^{\mu + 1} \: L_n^{(2\mu-1)}(2k) \: \expe^{-k} \: J_\mu(kz) \: \dif k 
 = \frac{r^{1/\alpha - 5/2}}{\left(1-t\right)^{2\mu}} \: \int_0^\infty k^{\mu+1} \: \expe^{-k(1+t)/(1-t)} \: J_\mu(kz) \: \dif k \\
& = \frac{2^{\mu+1} \: \Gamma(\mu + 3/2)}{\sqrt{\pi}} \: r^{1/\alpha - 2 + l} \: \frac{(1+t)(1-t)^2}{\left[(1+t)^2 + (1-t)^2z^2\right]^{\mu + 3/2}} \\
& = \frac{2^{\mu+1} \: \Gamma(\mu + 3/2)}{\sqrt{\pi}} \:
\frac{r^{1/\alpha-2+l}}{(1+r^{1/\alpha})^{3/2+\mu}} \: \sum_{n=0}^\infty t^n \: \left[C_n^{(\mu + 3/2)}(\xi) - C_{n-1}^{(\mu + 3/2)}(\xi) - C_{n-2}^{(\mu + 3/2)}(\xi) + C_{n-3}^{(\mu + 3/2)}(\xi)\right].
\end{align}
The last line can be simplified by adding together two Gegenbauer polynomial recursion relations \citep[Eqs~8.933(2) \& 8.933(3)]{gradshteyn4}, resulting in 
\begin{equation}\label{eq:rho_nl_def}
  \rho_{nl}(r) = \frac{A_{nlm}}{16\pi\alpha^2} \: \frac{2^{\mu+1}\Gamma(\mu + 1/2)}{\sqrt{\pi}}\frac{r^{1/\alpha-2+l}}{(1+r^{1/\alpha})^{3/2+\mu}}\left[\left(n+\mu+1/2\right)\:C_n^{(\mu+1/2)}(\xi)-\left(n+\mu-1/2\right)\:C_{n-1}^{(\mu + 1/2)}(\xi)\right]. 
\end{equation}
Just as for the potential basis functions, evaluating $n$ density basis functions requires only $O(n)$ steps. It is now straightforward to verify that the basis functions satisfy Eq.~(\ref{eq:basis_function_equation}). If the spherical harmonics are normalised to $J_{lm}=4\pi$ and we set $A_{nlm} = 1$, the overall normalisation constant for the basis functions written in Eqs.~\eqref{eq:phi_nl_def} and \eqref{eq:rho_nl_def} is
\begin{equation}
\label{eq:normalize}
N_{nl} = \frac{\Gamma(2\mu + n)}{2^{2\mu} \: n! \: 8\pi\alpha}.
\end{equation}
For numerical purposes, it may be desirable to redefine $A_{nlm}$ to incorporate more of the $n$ and $l$-dependent prefactors in Eqs. \eqref{eq:phi_nl_def} and \eqref{eq:rho_nl_def} (see Section~\ref{sec:implementation}).

\begin{figure}
  \centering
  \includegraphics{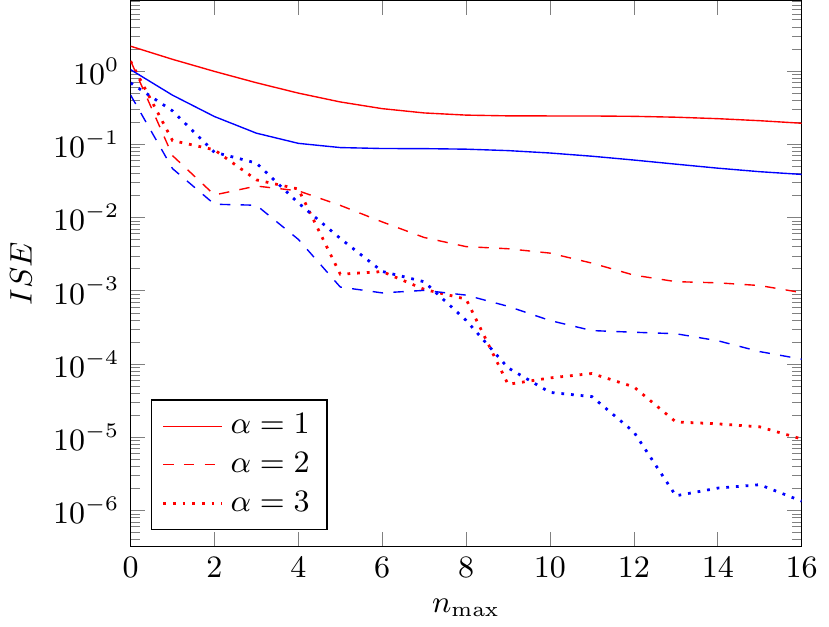}%
  \includegraphics{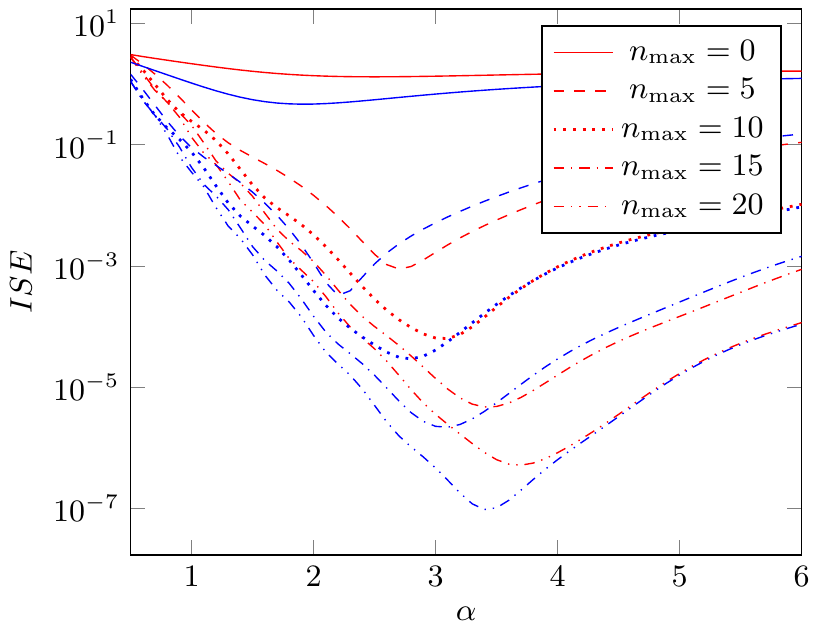}
  \caption{Left: The integrated squared error as defined in Eq.~\eqref{eq:ise} between the exact NFW profile and the profile reconstructed with radial basis functions up to order $n_\mathrm{max}$ for different values of $\alpha$ in Zhao's basis set (red lines) and the new basis set in this paper (blue lines). Right: The integrated squared error between the exact NFW profile and the profile reconstructed with radial basis functions, against the parameter $\alpha$ that characterises each basis set for different numbers of radial basis functions. Again red lines use Zhao's basis set and blue lines use the new basis set.}\label{fig:error}
\end{figure}

\begin{figure}
\centering
\includegraphics{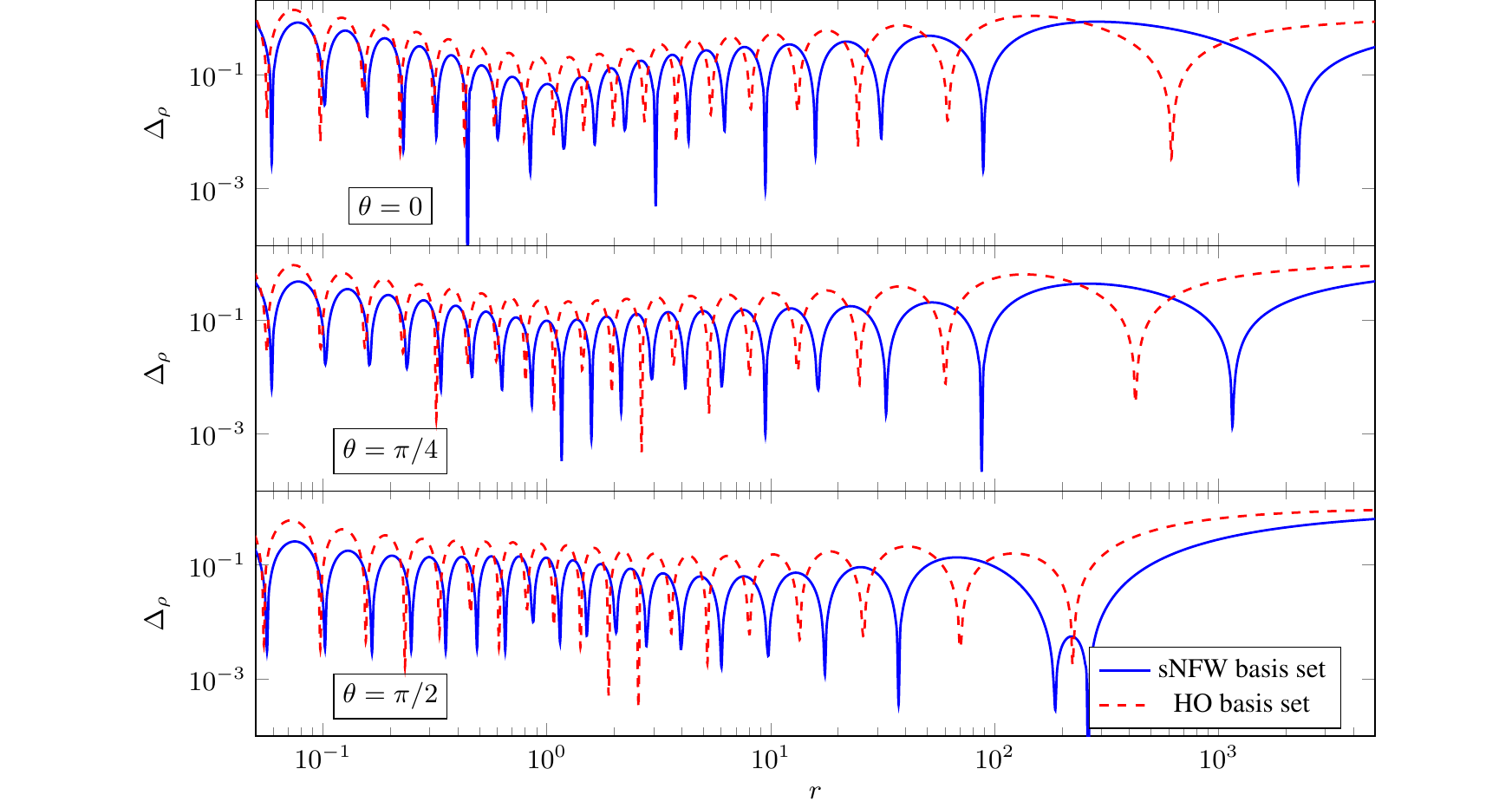}
\caption{Expansion of a flattened ($q = 0.8$) NFW density. Viewing angle $\theta$ is shown on each plot. Both expansions use radial terms up to $n=20$ and angular terms up to $l=12$. The error measure is
  $\Delta_\rho \equiv \log{\left|1 -
      \rho_\mathrm{approx}/\rho_\mathrm{exact}\right|}$ (lower is better; the dips are due to the oscillations around the true value).}\label{fig:flattened}
\end{figure}


\begin{figure}
\centering
\includegraphics{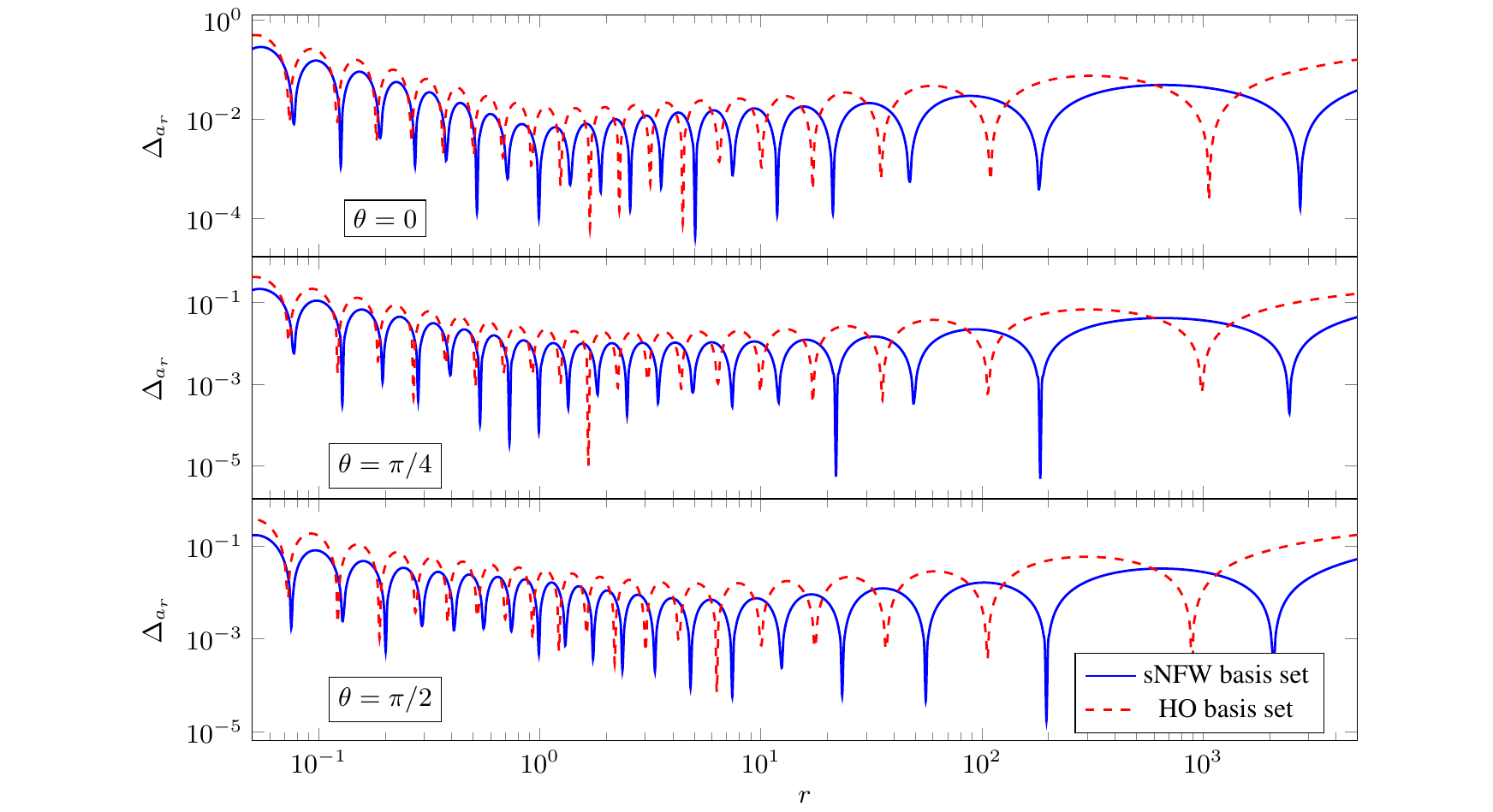}
\caption{Expansion of the radial acceleration in a flattened ($q = 0.8$) NFW potential. Viewing angle $\theta$ is shown on each plot. Both expansions use radial terms up to $n=20$ and angular terms up to $l=12$. The error measure is $\Delta_{a_r} \equiv \log{\left|1 - a_{r\:\mathrm{approx}}/a_{r\:\mathrm{exact}}\right|}$.}\label{fig:flattened_accel}
\end{figure}

\subsection{Comparison with the Zhao and Hernquist-Ostriker Basis Sets}\label{sec:zhao}
Prior to our work, only a single family of biorthogonal potential-density basis functions was known \citep[although][provided an alternate single set of solutions with no free parameters which our results now encompass]{rahmati2009}. This basis set was introduced by \cite{zhao1996}.  It is written in terms of a parameter $\alpha$ that affects both the inner and outer slope of the double-power-law.
\begin{equation}\label{eq:zhao}
\Phi_{nl}^{\mathrm{Zhao}} = \frac{r^l}{\left(1 + r^{1/\alpha}\right)^{\mu}} \: C^{(\mu + 1/2)}_n\left(\xi\right), \:\:\:\:\:\:\:\:
\rho_{nl}^{\mathrm{Zhao}} = \frac{r^{l - 2 + 1/\alpha}}{\left(1 + r^{1/\alpha}\right)^{\mu + 2}} \: C^{(\mu + 1/2)}_n\left(\xi\right)
\end{equation}
where $\mu$ and $\xi$ are defined as in Eq.~\eqref{eq:phi_nl_def}. The specific case when $\alpha =1/2$ was discovered by \citet{cluttonbrock1973} and has the \citet{Plummer1911} model as its lowest order term. The case when $\alpha =1$ was first proposed by \citet{hernquist1992} and is the most widely used of all the basis function expansions, with the \citet{Hernquist1990} model as its lowest-order term.

We note immediately the close similarity in form to our previous
definitions in \eqref{eq:phi_nl_def} and \eqref{eq:rho_nl_def}, where
the basis functions are also expressed in terms of the Gegenbauer
polynomials. Comparing the lowest-order density of the Zhao basis set
with that of our new basis set,
\begin{equation*}
\rho_{00} \propto \frac{1}{r^{2 - 1/\alpha}\left(1 + r^{1/\alpha}\right)^{\alpha + 3/2}}, \qquad\qquad\qquad
\rho_{00}^{\mathrm{Zhao}} \propto \frac{1}{r^{2 - 1/\alpha}\left(1 + r^{1/\alpha}\right)^{\alpha + 2}}.
\end{equation*}
Evidently, the only difference is the shallower outer slope in the
former case. This is significant as popular models for dark matter
haloes tend to have outer slopes closer to $r^{-3}$. For example, the
generalised NFW profile \citep{navarro2004} has $\rho \propto
r^{-\gamma}(1+r)^{\gamma-3}$, with values for the inner slope $\gamma$
ranging from $0.7$ to $1.5$, and outer slope fixed to $r^{-3}$.

In both our and Zhao's expansion families, when the inner slope is
fixed ($\gamma\equiv 2 - 1/\alpha$) the asymptotic outer slope is then
constrained. To give some examples, if we set $\gamma = 0.7$, then
Zhao's outer slope is $r^{-4.3}$ whereas ours is $r^{-3.65}$; and if
$\gamma = 1.5$ then Zhao's is $r^{-3.5}$ and ours is $r^{-3.25}$. We
therefore expect that our basis functions will be more efficient than
Zhao's for describing typical dark matter haloes, such as those
resimulated by~\citet{lowing2011}.

Before we pass on to numerical implementation of our new family, we
remark that our and Zhao's biorthonormal expansions do not by any
means exhaust all the possibilities. \rd{For example, we have
  identified a further family of basis expansions for which the
  zeroth-order members have density profiles that follow Einasto-like
  laws~\citep[c.f.,][]{Einasto65, Merritt06}. This will be the subject
  of a later paper.}

\section{Numerical Performance of the Basis Function Expansion}\label{sec::numerical}

We turn to the application of our new basis expansion to the representation of cosmological haloes. Before inspecting both analytic and numerical haloes in Sections \ref{Section::NFWhaloes} and ~\ref{Section::Numhaloes}, we present a formulation of the basis expansion that is computationally friendly. 

\subsection{Numerical Implementation}\label{sec:implementation}

It is very important to take advantage of a number of recursion relations for purposes of speed. It is also more efficient to factor out the constant parts of the potential-density pairs such that only the spatially-dependent pieces are calculated for each particle. We therefore modify the definition of the basis functions as
\begin{align}
{\hat\rho}_{nlm}(\vec{r}) = \hat{\rho}_{nl}(r)Y_{lm}(\theta,\phi) \qquad\qquad\qquad
&{\hat\rho}_{nl} = \frac{r^{1/\alpha-2+l}}{(1+r^{1/\alpha})^{\mu+3/2}}\Big((n+\mu+\tfrac{1}{2})C_n^{(\mu+1/2)}(\xi)-(n+\mu-\tfrac{1}{2})C_{n-1}^{(\mu+1/2)}(\xi)\Big),\\
{\hat\Phi}_{nlm}(\vec{r}) = \hat{\Phi}_{nl}(r)Y_{lm}(\theta,\phi)\qquad\qquad\qquad
&{\hat\Phi}_{nl} = \frac{\Beta_\chi(\mu,1/2)}{2\:r^{1+l}} - \frac{2 r^l}{(z^2+1)^{\mu + 1/2}} \: \sum_{j=0}^{n-1} \frac{j!\Gamma(2\mu)}{\Gamma(2\mu+j+1)} \: C_j^{(\mu + 1/2)}(\xi),
\end{align}
so that Poisson's equation becomes  $\nabla^2 {\hat \Phi}_{nlm}=4\pi K_{nl} {\hat \rho}_{nlm} $ with
\begin{equation}
K_{nl} = -\frac{n!\Gamma(2\mu)}{8\pi\alpha^2\Gamma(2\mu+n)}, 
\end{equation}
and the normalization constant
\begin{equation}
\hat{N}_{nlm} = \int\mathrm{d}^3\vec{r}\,\hat\Phi^{}_{nlm}(\vec{r})\hat\rho^*_{nlm}(\vec{r})=J_{lm}\frac{\alpha\sqrt{\pi}\Gamma(\mu)}{2^{1+2\mu}\Gamma(\tfrac{1}{2}+\mu)}, \qquad\qquad\qquad
J_{lm}=\int\mathrm{d}\phi\,\mathrm{d}\theta\sin\theta\,Y_{lm}(\theta,\phi) Y_{lm}^*(\theta,\phi).
\end{equation}
The Gegenbauer polynomials can be constructed recursively using the relation \citep[Eq.~8.933(1)]{gradshteyn4}
\begin{equation}
n C_n^{(w)}(\xi) = 2(w+n-1)\xi C_{n-1}^{(w)}(\xi)-(2w+n-2)C_{n-2}^{(w)}(\xi),
\end{equation}
where $C_0^{(w)}(\xi)=1$ and $C_1^{(w)}(\xi)=2w\xi$. To construct the potential function ${\hat \Phi}_{nl}$, we define
\begin{equation}
A_n=\frac{2n!\Gamma(2\mu)}{\Gamma(2\mu+n+1)},\\
\end{equation}
which satisfies the recurrence relation
\begin{equation}
\frac{A_{n+1}}{A_{n}}=\frac{n+1}{n+1+2\mu},\:\:\:\:\:\: A_0=1/\mu,
\end{equation}
such that the ladder of potential functions at fixed $l$ are given by
\begin{equation}
\hat\Phi_{nl}=\hat\Phi_{(n-1)l}-\frac{r^l}{(1+z^2)^{\mu+1/2}}A_{n-1}C_{n-1}^{(\mu+1/2)}(\xi).
\label{eq::pot_recursion}
\end{equation}
\rd{A naive implementation of this recursion relation, wherein one builds
up the higher-$n$ terms by starting from $\Phi_{0l}$, results in large
errors for high $n$. In the left panel of
Fig.~\ref{fig::numerical_accuracy}, we show the logarithm of the
relative difference between the potential computed using this naive
approach and an arbitrary precision calculation from
\emph{Mathematica}. The error grows with increasing $n$ and $l$, as
the computation requires taking a small difference between large
numbers. To see this we note that a valid series expansion of the
incomplete beta function is (using \citet[8.17.8]{dlmf} and
\citet[Eq.~2.5]{Fi61}) 
\begin{equation}
\Beta_\chi(\mu,1/2) =
\frac{\chi^\mu}{\sqrt{1+z^2}} \sum_{j=0}^\infty A_j
C_j^{(\mu+1/2)}(\xi).
\end{equation}
Comparison with \eqref{eq::pot_recursion} shows
that the terms in $\hat\Phi_{nl}$ tend to zero as $n\to\infty$,
resulting in the aforementioned catastrophic cancellation. This
expression, however, provides us with an alternative method of
computing $\hat\Phi_{nl}$. At some large order $N$,
e.g. $N=2n_\mathrm{max}$, we assume that $A_N \approx 0$. Then we can
write
\begin{equation}
\hat\Phi_{nl} \approx \frac{r^l}{(1+z^2)^{\mu+1/2}}\sum_{j=n}^{N}A_jC_j^{(\mu+1/2)}(\xi),
\end{equation}
which requires us to calculate $C^{(\mu+1/2)}_{n}(\xi)$ and $A_n$
recursively up to order $N$. We then recursively construct the potential
basis functions $\hat\Phi_{nl}$ downwards using
Eq.~\eqref{eq::pot_recursion} where now all $\hat{\Phi}_{nl}$ are
accurate to the magnitude of $A_{N}$. This
procedure~\footnote{\rd{It is analogous to Miller's method in
    numerical analysis, apparently introduced by J.C.P. Miller for the
    computation of Bessel functions~\citep[see e.g.,][]{AS}. }}
results in the reduced errors shown in the right panel of
Fig.~\ref{fig::numerical_accuracy}.  We only perform this procedure
for $l>4$ as for $l\leq4$ the naive implementation is satisfactory and
the decay of $\hat{\Phi}_{nl}$ is weak for small $l$.}

Finally, note that the $\hat{N}_{nl}$ are independent of $n$ and the $K_{nl}$ satisfy the properties
\begin{equation}
\frac{K_{(n+1)l}}{K_{nl}}=\frac{n+1}{n+2\mu},\:\:\:\:\:\: K_{0l}=-\frac{1}{8\pi\alpha^2}.
\end{equation}
With these definitions, a set of coefficients $\hat{C}_{nlm}$ are computed from a cloud of particles, and the potential and density reconstructed as
\begin{equation}
\hat{C}_{nlm} = \frac{1}{\hat{N}_{nlm}K_{nl}}\sum_i m_i \hat{\Phi}_{nl}(r_i)Y_{lm}(\theta_i,\phi_i),\qquad\qquad\qquad
\Phi = \sum_{nlm}\hat{C}_{nlm}\hat{\Phi}_{nlm},\qquad\qquad\qquad
\rho = \sum_{nlm}K_{nl}\hat{C}_{nlm}\hat{\rho}_{nlm}.
\end{equation}

\begin{figure}
\phantom{.}\hspace{0.185\paperwidth}{\large Simulation}\hspace{0.29\paperwidth}{\large Reconstruction}\\[6pt]
\phantom{.}\hspace{0.07\paperwidth}$x$--$z$ plane\hspace{0.07\paperwidth}$x$--$y$ plane\hspace{0.07\paperwidth}$y$--$z$ plane%
\hspace{0.07\paperwidth}$x$--$z$ plane\hspace{0.07\paperwidth}$x$--$y$ plane\hspace{0.07\paperwidth}$y$--$z$ plane\\[-6pt]
\includegraphics{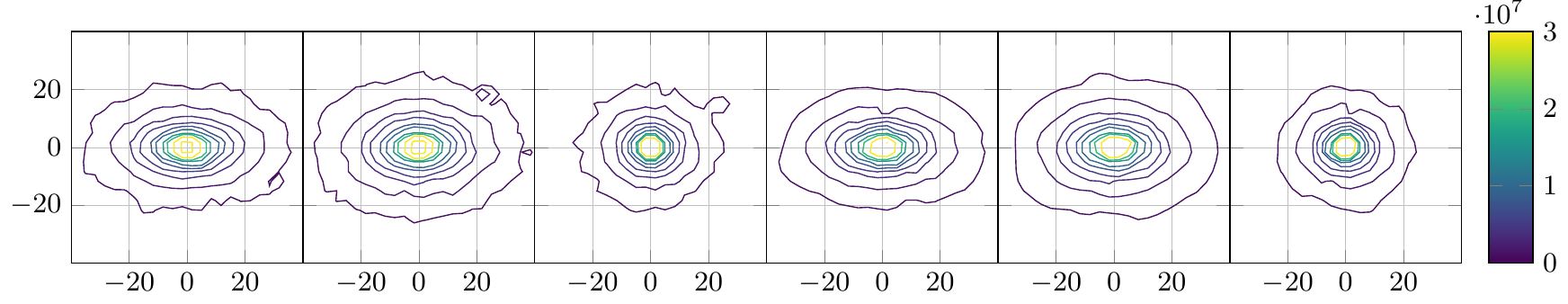}
\includegraphics{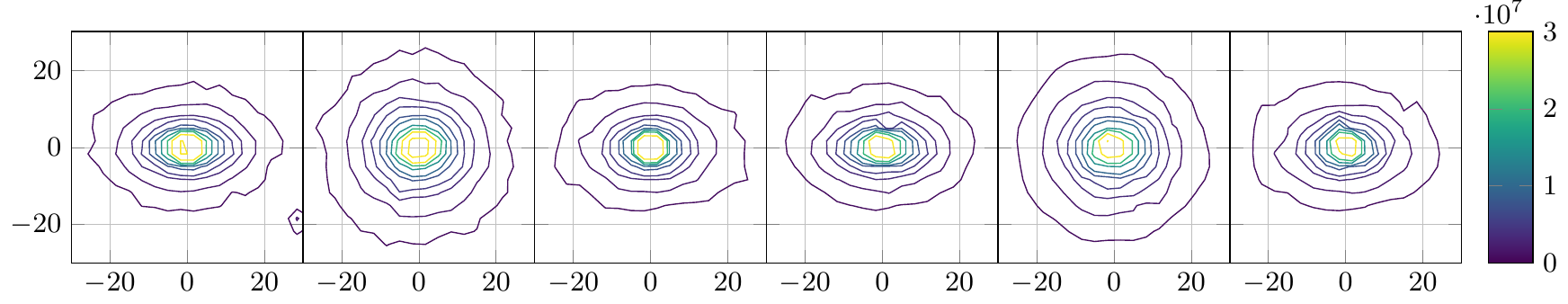}
\includegraphics{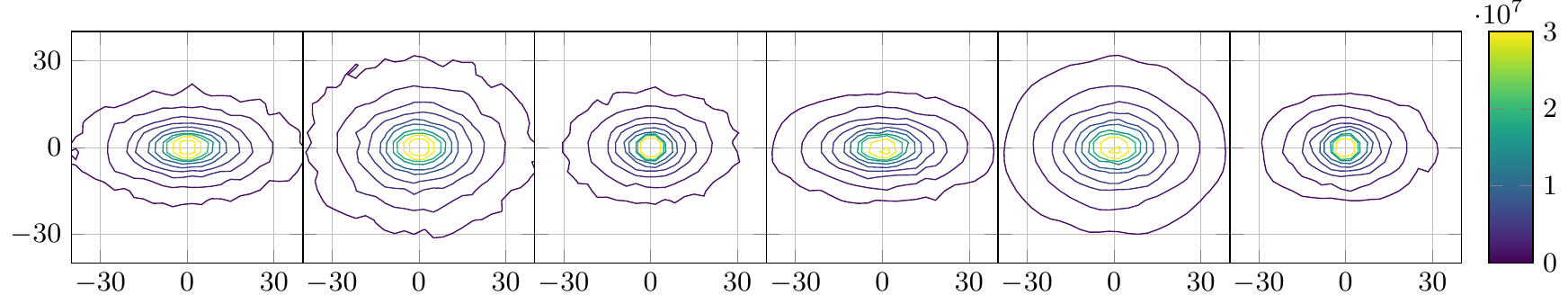}
\includegraphics{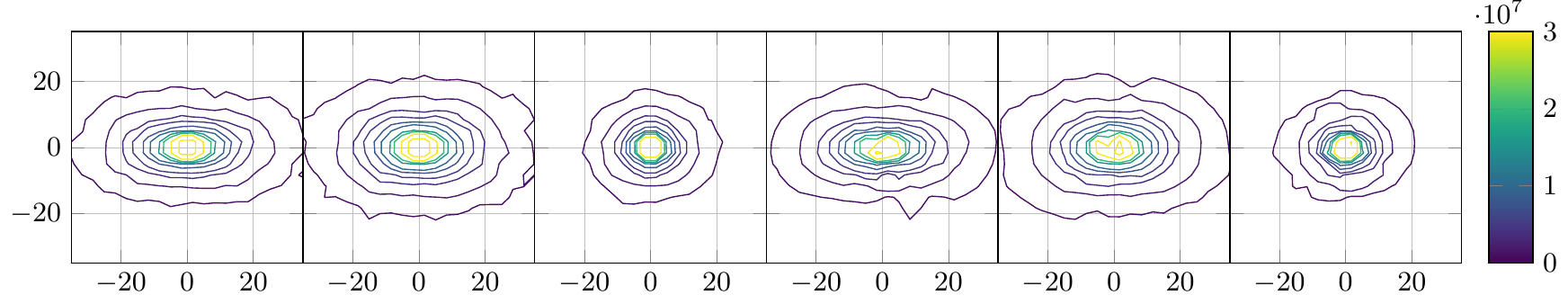}
\includegraphics{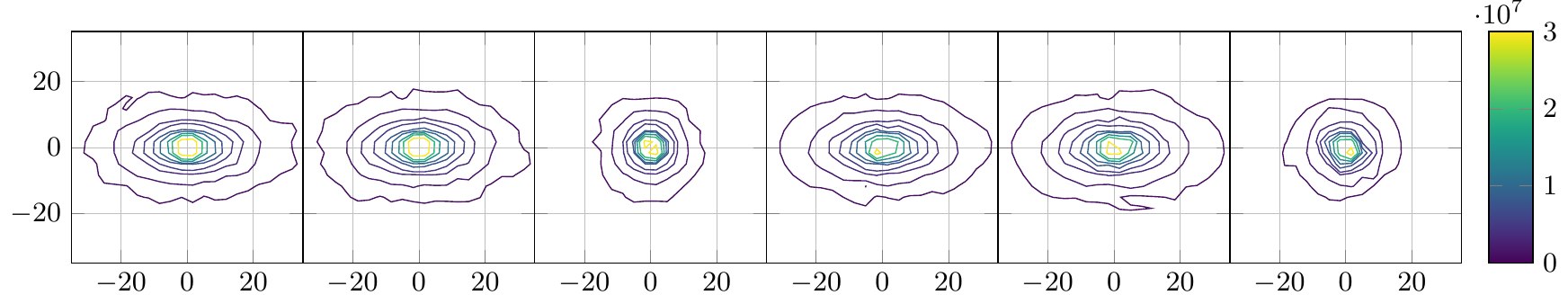}
\includegraphics{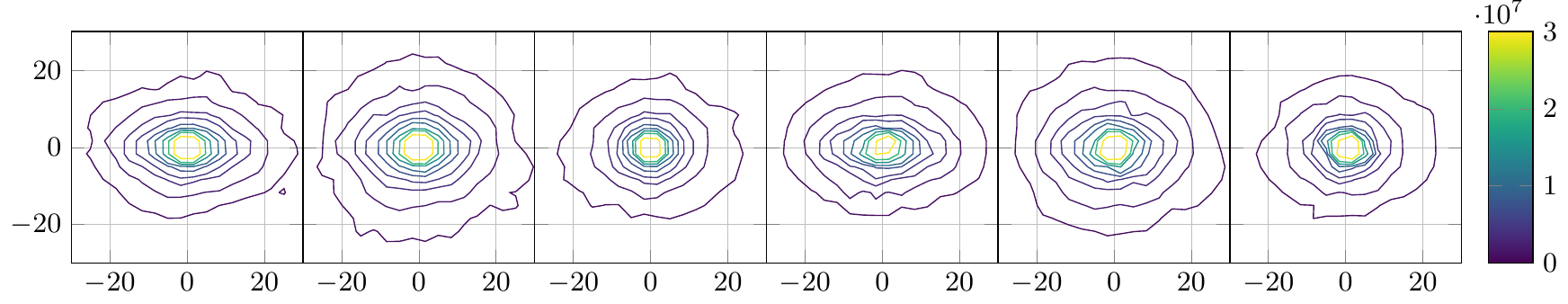}
\caption{From left to right: Density contours in the three principal planes of the original numerical halo, then density contours in the same three principal planes of the halo as reconstructed with basis functions up to $n = 20$ and $l = 12$. Distances are in kpc, and the densities are in $M_\odot\mathrm{kpc}^{-3}$.}\label{fig:halo_contours}
\end{figure}

\begin{figure}
\phantom{.}\hspace{0.185\paperwidth}{\large Simulation}\hspace{0.29\paperwidth}{\large Reconstruction}\\[6pt]
\phantom{.}\hspace{0.07\paperwidth}$x$--$z$ plane\hspace{0.07\paperwidth}$x$--$y$ plane\hspace{0.07\paperwidth}$y$--$z$ plane%
\hspace{0.07\paperwidth}$x$--$z$ plane\hspace{0.07\paperwidth}$x$--$y$ plane\hspace{0.07\paperwidth}$y$--$z$ plane\\[-6pt]
\includegraphics{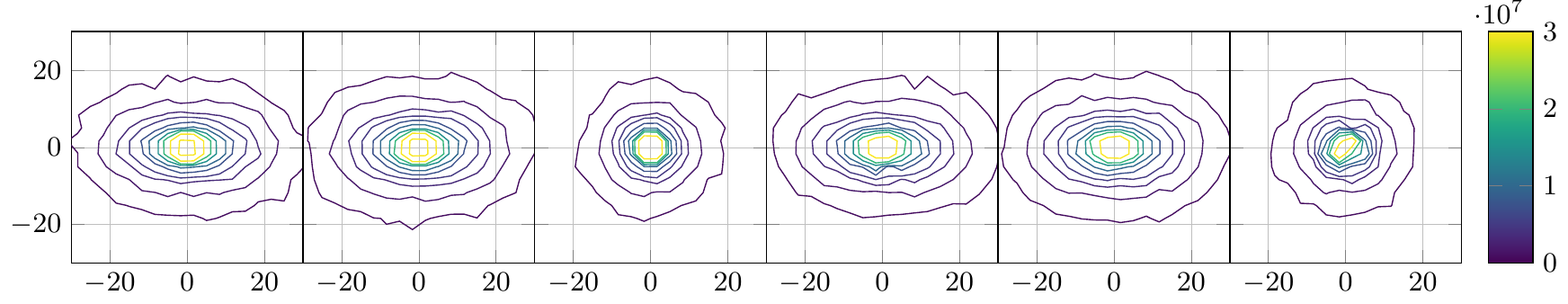}
\includegraphics{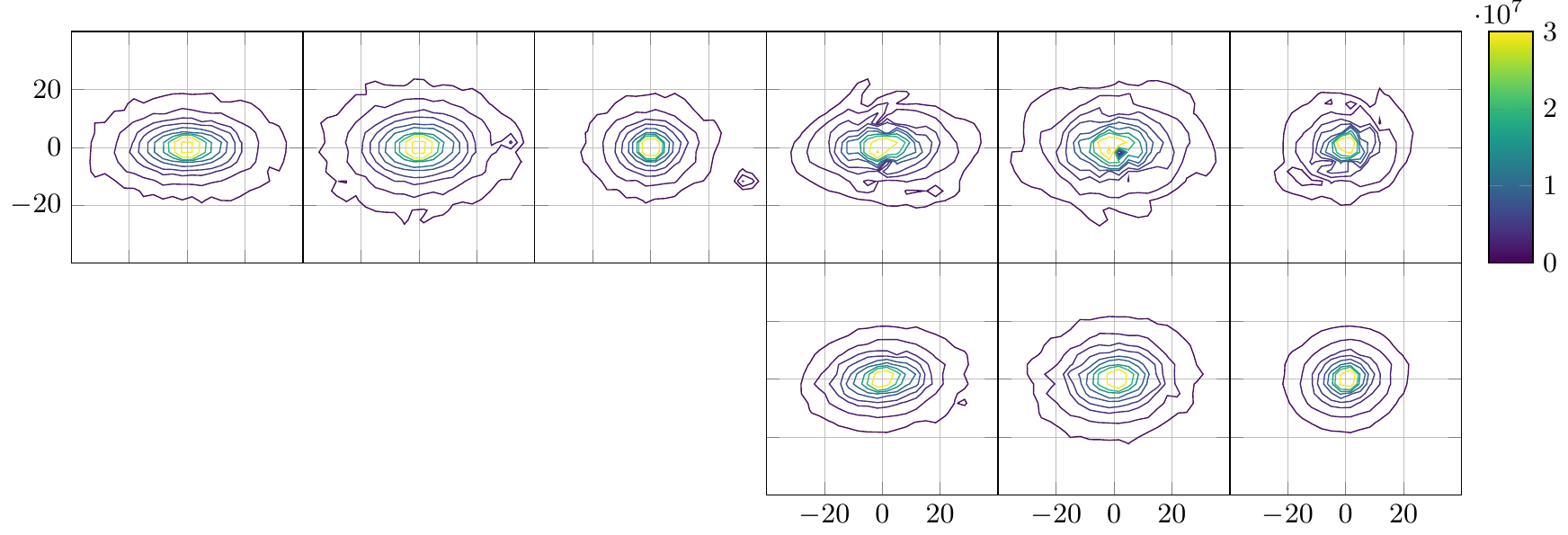}
\includegraphics{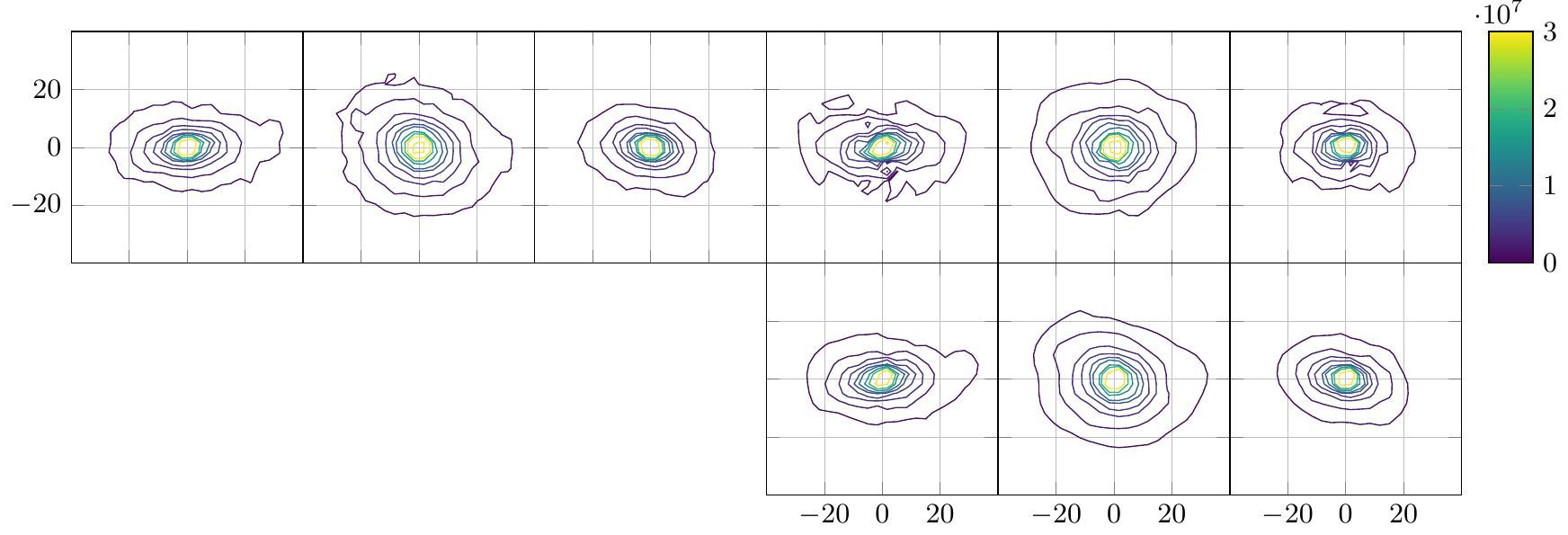}
\caption{Continuation of Fig.~\ref{fig:halo_contours}. The latter two
  haloes possess a large number of small satellites, so for these we
  display on the right two reconstructions: \rd{top, a reconstruction of the full halo;
  and bottom, one where the particles energetically 
  bound to the 50 most massive satellites have been removed.}
  This illustrates the issues
  surrounding expansion accuracy in the presence of
  sub-structure.
}\label{fig:halo_contours2}
\end{figure}

\subsection{Analytical haloes}\label{Section::NFWhaloes}

\subsubsection{Spherical NFW models}

We first consider the reconstruction of spherical NFW haloes using the radial terms of the expansion. We compare the $\alpha = 1$ member of our family of expansions with the \citet{hernquist1992} (hereafter HO) expansion, which is the $\alpha = 1$ member of the \citet{zhao1996} family. Both basis sets have lowest-order densities with $1/r$ cusps as $r\to 0$. In Fig.~\ref{fig:spherical}, we show the expansion of a spherical NFW potential and density, using only radial terms ($n \geq 0$, $l = 0$). With an equal number of terms, our $\alpha=1$ basis set performs better than the corresponding Hernquist-Ostriker set due to its closer approximation to the NFW profile in the asymptotic fall-off of the lowest-order density, $\rho \sim r^{-7/2}$. The corresponding behaviour of the Hernquist-Ostriker basis set is $\rho \sim r^{-4}$. Note that the $r$-axis of Fig.~\ref{fig:spherical} is logarithmically scaled and measured in units of the scalelength, so we in fact get several hundred additional scalelengths of accurate behaviour. The improved convergence at large radii also reduces the amplitudes of the oscillations at smaller radii. The expansion of the potential is always more accurate than that of the density because the oscillations are integrated over, and therefore effectively smoothed; the accuracy of the radial acceleration expansion lies between that of the potential and the density. According to Fig.~\ref{fig:coeffs}, the coefficients used in the spherical NFW expansion empirically follow power-laws for both basis sets. In our expansion, $C_{n00} \sim n^{-2}$ whereas in the Hernquist-Ostriker expansion $C_{n00} \sim n^{-3}$.

To make quantitative statements about the error, we follow
\citet{Vasiliev2013} and calculate the \textit{integrated squared
  error}. This is the mass-weighted fractional density error defined
by
\begin{equation}\label{eq:ise}
ISE = \int_{r > r_\mathrm{min} \atop r < r_\mathrm{max}} \frac{\left|\rho_\mathrm{exact} - \rho_\mathrm{approx}\right|^2}{\rho_\mathrm{exact}} \: \dif^{\:3}\vec{r}.
\end{equation}
Here, $\rho_\mathrm{approx}$ is the density reconstructed using basis
functions up to a radial order of $n_\mathrm{max}$. \rd{Although other
  measures of error can be constructed, Vasiliev's suggestion is
  appealing as it is mass-weighted and does not bias the result
  towards the outer parts of the model.}  We use $r_\mathrm{min} =
0.01$ and $r_\mathrm{max} = 100$, as this is the range over which we
expect the expansions to be most applicable in astrophysical
problems. The run of this error measure with $n_\mathrm{max}$ is shown
on the left in Fig.~\ref{fig:error} for the two spherical NFW
expansions under consideration. It is clear both that higher values of
$\alpha$ give better accuracy, and that for given $\alpha$ and
$n_\mathrm{max}$ our basis set is more accurate than Zhao's in the
task of reproducing NFW haloes. To illustrate this difference, on the
right in Fig.~\ref{fig:error} we show how the optimum $\alpha$ (the
minimum of each plot) varies with differing $n_\mathrm{max}$, with our
basis set performing better than Zhao's in each and every case.

A detailed error analysis of basis function techniques is carried out in \citet{kalapotharakos2008}, who claim that a significant factor in obtaining high accuracy is choosing a basis set whose lowest-order density function has the correct inner slope behaviour. For a spherical NFW model, this would imply that the  $\alpha =1$ expansion is best, as the zeroth-order model behave like $\rho \sim r^{-1}$ at small radii. It is possible to choose combinations of the density basis functions so that the most diverging terms at the centre cancel, as \cite{hernquist1992} already showed by building a cored density model from the $n=0$ and $n=1$ monopole terms of their basis set. This delicate balance though is susceptible to numerical error. To reproduce accurately the precession of orbits that pass close to the centre, the criterion of \citet{kalapotharakos2008} seems reasonable. However, it is at odds with the plots in Fig.~\ref{fig:error}, which show that models with $\alpha \approx 3$ or $\rho \sim r^{-5/3}$ provide the smallest integrated square error in the density, if $n_\mathrm{max} \approx 20$. A similar result was obtained by \citet{Vasiliev2013}, who used an identical integrated error measure to the present work and concluded that a somewhat higher value of $\alpha$ is preferable for providing a global fit to cosmological haloes, even at the expense of accuracy of the inner slope exponent near the centre.  Clearly, the best choice depends on the application in hand. 

\subsubsection{Flattened NFW models}

It is important to consider flattened objects, as the haloes found in $N$-body simulations are generically flattened or triaxial \citep[e.g.,][]{Jing2002,Allgood2006}. We test the performance by attempting to reconstruct a flattened density profile and including the angular terms or $(l,m)$ terms in the series. An axisymmetric NFW density profile is $\rho = {\bar m}^{-1}(1+ \bar{m})^{-2}$, where ${\bar m}^2 = x^2 + y^2 + z^2/q^2$. This means the density is stratified on similar concentric spheroids with axis ratio $q$. Figs \ref{fig:flattened} and \ref{fig:flattened_accel} shows the expansion of a flattened ($q = 0.8$) NFW density, and the corresponding acceleration due to the potential. In each case,  we compare the $\alpha = 1$ member of our family of expansions with the Hernquist-Ostriker expansion, which is the $\alpha = 1$ member of the \citet{zhao1996} family. Typically, we use $n_{\rm max} =20$ radial basis functions and $l_{\rm max} =12$ angular basis functions.

Each of the reconstructed quantities is plotted along three polar angles ($\theta$). Note that the convergence is always superior nearer the equator ($\theta = \pi/2$) than at the poles ($\theta = 0$). This is a feature of any expansion involving spherical harmonics and can be remedied by introducing additional `$l$' terms. In the flattened case, both expansions lose accuracy in the very inner  and outer parts of the haloes. As the $l$-dependence of both our $\alpha=1$ set and the HO set is similar, we do not expect either basis set to be favoured in this regard. However, the superior behaviour in the outskirts of our basis set is maintained. \citet{lowing2011} used the HO basis set to represent haloes, where on the order of tens of terms are used in both the angular and radial directions. They found errors of $<10\%$ are achieved over a few hundred kiloparsecs; we expect our basis set to provide improved accuracy in this regime.

\subsection{Numerical Haloes}\label{Section::Numhaloes}

We now analyse a collection of ten Milky Way-like dark matter haloes, extracted from a suite of cosmological $N$-body zoom-in simulations. These simulations are run with the $N$-body part of \textsc{gadget-3} which is similar to \textsc{gadget-2} \citep{springel_2005}. The zoom-in strategy follows \cite{onorbe} and all initial conditions are generated with \textsc{music} \citep{music}. Cosmological parameters are taken from the \cite{planck} with $h=0.679$, $\Omega_b = 0.0481$, $\Omega_0=0.306$, $\Omega_\Lambda=0.694$, $\sigma_8=0.827$, and $n_s=0.962$. In order to select our haloes, we first simulate a 50$h^{-1}$ Mpc box with $512^3$ particles from $z=50$ to $z=0$. We use  \textsc{rockstar} \citep{rockstar} to identify haloes and select Milky Way-like haloes which have virial masses between $7.5\times10^{11}M_\odot-2\times10^{12} M_\odot$, no major mergers since $z=1$, and no haloes with half the mass of Milky Way analogue's mass within $2h^{-1}$ Mpc. For each selected halo, we select all particles within 10 virial radii and run an intermediate resolution zoom-in whose maximum resolution is $2048^3$, corresponding to a particle mass of $1.8\times10^6 M_\odot$. This intermediate step helps to reduce the contamination from low resolution particles in our final, high resolution zoom-in. For the final zoom-in, we take the intermediate resolution simulation and select all particles within 7.5 virial radii. We then run a zoom-in with a maximum resolution of $4096^3$, corresponding to a particle mass of $2.23\times10^5 M_\odot$. All of our high resolution zoom-ins are uncontaminated within $1h^{-1}$ Mpc of the main halo. The detailed properties of each halo are given in Table~\ref{tbl:halo_props}.

\begin{table}
\centering
\begin{tabular}{ccccccccc} \hline
  $N / 10^6$ & $M_v / (10^{12} M_\odot)$ & $r_v / \mathrm{kpc}$ & $r_s / \mathrm{kpc}$ & $r_\mathrm{iso} / \mathrm{kpc}$        & $\alpha$ & $p$   & $q$   & Figure \\ \hline 
  10.3       & 1.88                     & 325                  & 65.2                & 33.6                & 1.22     & 0.860 & 0.811 & \ref{fig:halo_contours} \\ 
  5.0        & 0.91                     & 256                  & 37.8                & 20.5                & 1.18     & 0.972 & 0.816 & \ref{fig:halo_contours} \\ 
  8.7        & 1.57                     & 306                  & 47.6                & 25.8                & 1.18      & 0.889 & 0.761 & \ref{fig:halo_contours} \\ 
  7.4        & 1.36                     & 292                  & 52.8                & 25.9                & 1.26      & 0.800 & 0.733 & \ref{fig:halo_contours} \\ 
  5.9        & 1.07                     & 269                  & 58.1                & 30.7                & 1.20      & 0.804 & 0.795 & \ref{fig:halo_contours} \\ 
  5.5        & 0.89                     & 254                  & 46.0                & 23.7                & 1.22      & 0.909 & 0.834 & \ref{fig:halo_contours} \\ 
  4.9        & 0.89                     & 253                  & 35.7                & 19.8                & 1.16      & 0.807 & 0.780 & \ref{fig:halo_contours2} \\ 
  11.3       & 1.62                     & 310                  & 123                 & 38.1                & 1.59      & 0.858 & 0.769 & \ref{fig:halo_contours2} \\ 
  7.7        & 1.11                     & 273                  & 98.5                & 32.9                & 1.54      & 0.926 & 0.779 & \ref{fig:halo_contours2} \\ 
  7.6        & 1.71                     & 315                  & 132.6               & 47.6                & 1.49     & 0.861 & 0.730  & \ref{fig:halo_contours3} \\ 
\hline
\end{tabular}
\caption{The properties of each numerical halo in our sample. The haloes are listed in the order that they are displayed in the figures. $N$ is the number of particles, $M_v$ is the virial mass, $r_v$ is the virial radius, $r_s$ is the scalelength, $r_\mathrm{iso}$ is the distance at which the slope is isothermal (so $r_{\rm iso} = r_s/(1/2 + \alpha)^\alpha$), $\alpha$ parameterises the inner and outer slopes of the density basis functions, $p$ is the $y$--$x$ axis ratio, and $q$ is the $z$--$x$ axis ratio.}
\label{tbl:halo_props}
\end{table}

For each halo, we take all of the particles within 500 kpc of the main halo in $z=0$ snapshot. This corresponds to between 5-12 million particles, depending on the halo mass.  We wish to investigate the ability of our new basis sets to represent these numerical haloes. To this end, we must first choose the two global parameters that specify the basis set -- the scalelength $r_s$ and the parameter $\alpha$. We need not worry at this stage about the overall normalisation (related to the total mass) because this will be set automatically when performing the sum over particles via Eq.~\eqref{eq:particle_sum}. To set these, we first fit the zeroth-order density function $\rho_{000}$, which is a spherically-symmetric model. Because we know the virial radius $r_v$ for each halo, we perform the fitting procedure in terms of the dimensionless \textit{concentration} $c \equiv r_v/r_s$, as is standard in the literature. The particles are binned logarithmically in radius, and a non-linear least squares algorithm then adjusts $\alpha$ and $c$ to minimise the difference between the logarithms of the inferred bin density and the model density. \rd{For this fitting procedure, we use a form of the zeroth-order density function that is parametrised by the mass enclosed by the virial radius, i.e. that satisfies $\int_0^{r_v} 4\pi \rho(r) r^2 \dif r = M_v$. This is
\begin{equation}\label{eq:zeroth_density_virial}
  \rho(r)  = \left[\Beta_{\chi_v}(\alpha,1/2) - \frac{c/\alpha}{\left(1 + c^{1/\alpha}\right)^{\alpha+1/2}}\right]^{-1} \: \frac{M_v \: (\alpha + 1/2)}{4 \pi \alpha^2 \: (r_v/c)^3} \: \frac{(cr/r_v)^{1/\alpha - 2}}{\left(1 + (cr/r_v)^{1/\alpha}\right)^{\alpha + 3/2}},
\end{equation}
where $\chi_v \equiv c^{1/\alpha}/\left(1 + c^{1/\alpha}\right)$.}

\begin{figure}
\phantom{.}\hspace{0.25\paperwidth}$x$--$z$ plane\hspace{0.08\paperwidth}$x$--$y$ plane\hspace{0.08\paperwidth}$y$--$z$ plane\\[-14pt]
\begin{center}
\includegraphics{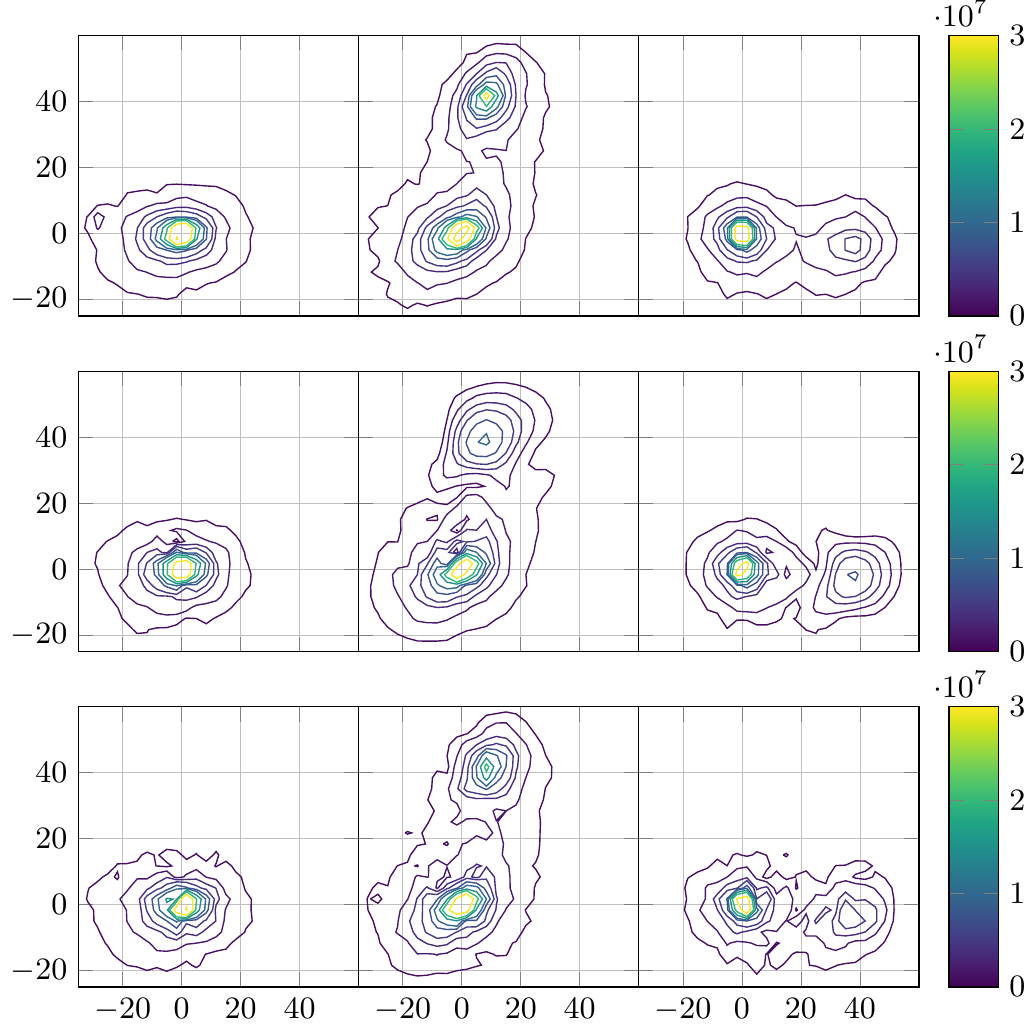}
\end{center}
\caption{A halo with a prominent, massive satellite. From top to bottom: the original halo; the reconstruction with maximum order $n = 20$ and $l = 12$; the same with $n = 40$ and $l = 40$.}\label{fig:halo_contours3}
\end{figure}

We can now perform a full expansion, using for each halo the basis set with the determined best values of $\alpha$ and $r_s$. To compare the accuracy in the reproduction of the density, we draw contour plots in each principal plane, as shown in Figs \ref{fig:halo_contours}, \ref{fig:halo_contours2} and \ref{fig:halo_contours3}. The smooth underlying density distribution in each principal plane of the original numerical halo is estimated by taking particles in a slab of width $2 \: \mathrm{kpc}$ around the plane, then creating a two-dimensional histogram with each bin having an area of approximately $3 \: \mathrm{kpc}^2$. The density as reconstructed from the basis function expansion is sampled along rays in each principal plane, and then interpolated onto a grid. The same 12 contours are drawn on each plot, spaced approximately logarithmically between $10^6$ and $2\times 10^8$ $M_\odot \: \mathrm{kpc}^{-3}$.

\begin{figure}
  \centering
  \includegraphics{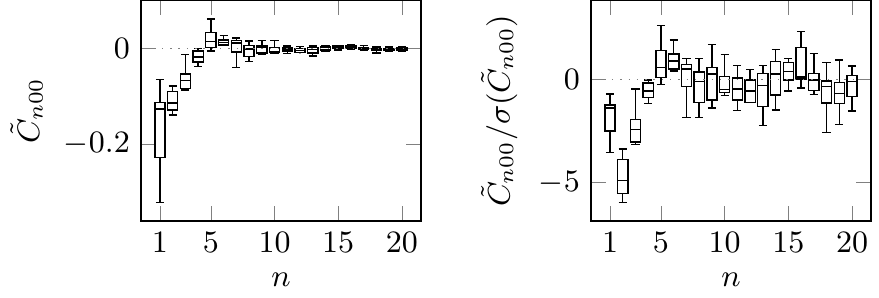}%
  \hfill%
  \includegraphics{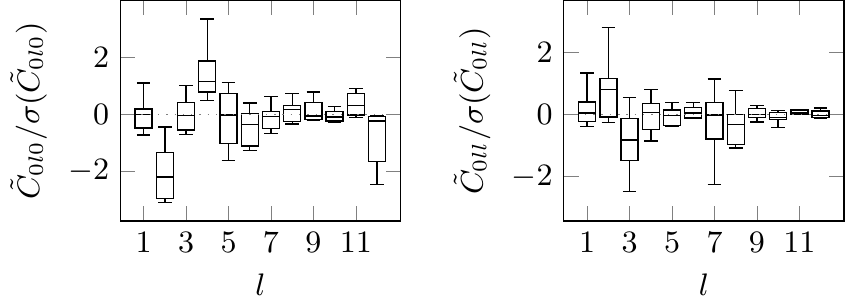}%
  \caption{We denote normalised coefficients by tildes ($\tilde{C}_{nlm} \equiv C_{nlm}/C_{000}$) and the standard deviation of a variable $x$ by $\sigma(x)$. Panels 1 and 2: the radial coefficients $\tilde{C}_{n00}$ for nine numerical haloes. Panel 3: the angular coefficients $\tilde{C}_{0l0}$. Panel 4: the angular coefficients $\tilde{C}_{0ll}$. The latter three plots are normalised by each coefficient's standard deviation to make details easier to see. The trend in the  $C_{0l0}$ coefficients indicates the flattening of the haloes along the $z$-axis, and the trend in the $\tilde{C}_{0ll}$ coefficients shows the elongation along the $x$-axis.}\label{fig:boxplot_n}
\end{figure}

The haloes displayed in Figs \ref{fig:halo_contours} and
\ref{fig:halo_contours2} are rotated such that the principal planes are aligned with the coordinate axes (shortest axis along the $z$-axis), so that the angular expansion coefficients can be compared meaningfully. Distributions of the expansion coefficients for these nine numerical haloes are shown in Fig.~\ref{fig:boxplot_n}. One could produce an artificial `halo' with geometry typical for this family of nine numerical haloes by drawing coefficients from the distributions shown in these plots. From Figs \ref{fig:halo_contours} and \ref{fig:halo_contours2}, we see that key features of the haloes are resolved correctly, including:
\begin{inparaenum}
\item the orientation of the principal axes,
\item the axis ratios in the three principal planes, and
\item the run of ellipticity with radius.
\end{inparaenum}
The size of the smallest resolvable feature is limited by the distance between the roots of the polynomial used to define the highest-order function used in the reconstruction. 

\rd{Two haloes in Fig.~\ref{fig:halo_contours2} demonstrate one pitfall of the method. The presence of unresolved but massive sub-haloes can cause a blow-up in the higher-order coefficients of the expansion. This would be cancelled by yet higher-order terms, but as the expansion is truncated these are not present. This problem is generic -- it affects other basis sets, such as Zhao's, to a greater or lesser degree -- but it can be remedied by removing the unresolved sub-haloes by an automated halo-finding procedure, as demonstrated in the figure. When these sub-haloes are removed the `optimal' values of $\alpha$ and $r_s$ (as determined by fitting Eq.~\eqref{eq:zeroth_density_virial}) are reduced to values more characteristic of the other haloes in the sample.}

The final halo, displayed in Fig.~\ref{fig:halo_contours3}, provides a serious challenge. It is accompanied by a massive close-in satellite with a mass of $1.9\times10^{11} M_\odot$, and thus has a highly aspherical geometry. We therefore examine this halo in greater detail in Fig.~\ref{fig:halo_contours3} using $n_{\rm max} = 20$ and $l_{\rm max} = 12$ (middle panel) and $n_{\rm max} = 40$ and $l_{\rm max} = 40$ (lower panel). Remarkably, the overall structure of both the halo and the large satellite are correctly resolved even with $n_{\rm max} =20$. This is impressive, as we might have suspected at the outset that two basis function expansions, centered on each object, would be necessary to reproduce the merging structure.


\section{Conclusions}
\label{sec:conclusions}

Biorthogonal density and potential basis functions provide useful and flexible ways of describing realistic dark matter haloes and galaxies, which may be apsherical, triaxial or further misshapen. 
The coefficients of these basis-function expansions can be found
easily by summing over the particles in an $N$-body realisation and
used to reconstruct both the potential and density.  We have
discovered a completely new family of biorthogonal potential-density
pairs, parametrised in terms of $\alpha$. The zeroth-order model has a
density $\rho \sim r^{-2 + 1/\alpha}$ at small radii and $\rho \sim
r^{-3 -1/(2\alpha)}$ at large radii. This double-power law profile has
a central logarithmic slope between $0$ and $-2$ and an asymptotic
logarithmic slope between $-3$ and $-4$, making it perfect for
representing dark haloes. The zeroth-order model with $\alpha =1/2$
has a harmonic core and is the celebrated perfect sphere of
\citet{dezeeuw1985}, whilst the model with $\alpha =1$ has a $1/r$
central density cusp and is the \rd{super-NFW}
model~\citep{Lilley2017b}. For each of these zeroth-order models, we
provide a biorthogonal basis function expansion in terms of standard
special functions readily available in numerical libraries. This
extends the zeroth-order model into the highly realistic regime of
flattened, distorted and triaxial density profiles.

Previously, the only known family of biorthogonal potential-density pairs was the one outlined by \citet{zhao1996}, of which the most widely-used member is the \citet{hernquist1992} expansion. The zeroth-order model has a density $\rho \sim r^{-2 + 1/\alpha}$ at small radii and $\rho \sim r^{-3 -1/\alpha}$ at large radii. Although the central density cusp is the same, the outer density falls off rather more quickly than in our expansion, making Zhao's family less well-matched to modelling dark haloes.

We have demonstrated the capabilities of our basis function expansions by using them to recreate spherical and flattened dark haloes with analytic densities of \citet*{NFW1997} form. In particular, we showed that our method represents a significant improvement over the \citet{hernquist1992} expansion, giving a more accurate reproduction of the density, potential and radial force of NFW-like models. Additionally, we decomposed 10 simulated cosmological haloes using our basis functions, computing the coefficients as simple sums over the particles. This yielded very encouraging results with highly flattened and triaxial dark halo density distributions well-reproduced with typically 20 radial and 12 angular terms (giving a total of $20\times12^2\approx3000$ terms). Simulated dark haloes can be lopsided, distorted or twisted, especially if there is a nearby companion exerting strong tidal forces (c.f. the Milky Way and the Magellanic Clouds). Particular striking is the ability of our basis function expansion to reproduce the density of a highly asymmetric dark halo which is in the process of accreting a companion large subhalo.

\citet{lowing2011} used the \citet{hernquist1992} expansion to create potentials approximating the structure of the dark haloes in the Aquarius simulations. They found that the radial force from the expansion sometimes differed from that in the simulations when the central logarithmic slope of the numerical halo density was shallower than the slope of $-1$ of the zeroth-order model.
They also highlighted a new application for biorthogonal basis function expansions as a tool for data compression of large and expensive $N$-body simulations. By calculating sets of coefficients from numerical snapshots at different redshifts and interpolating between them, they obtained a compact and efficient summary of the halo's evolution throughout time. This is very powerful because new objects can then be inserted and the entire simulation can be re-run cheaply making the method well-suited to study a range of problems, such as the dynamics of small satellites, the stripping of globular clusters or the shape of tidal streams.

Another important area of application is to the fitting of data on the Milky Way galaxy, which has increased substantially in both quality and quantity over the last few years. Models of the Milky Way galaxy are assembled from simple building blocks. Results of calculations using such `pieces of Lego' are often very troubling. For example, when the position and velocity data of stars in the Sagittarius stream are fitted to such models, the conclusion is that the dark halo is triaxial with the short and long axes in the Galactic plane~\citep{Law2010}. This configuration is unstable~\citep[see e.g.,][]{Debattista2013} and in conflict with observational data on the disk \citep{Kuijken1994}. The strong suspicion is that the inflexible model of the Milky Way's potential prevented proper exploration of parameter space and artificially confined the solution for the dark matter distribution into an unrealistic straitjacket. A completely new way of representing the Galactic dark halo is needed with the advent of large-scale photometric, spectroscopic and astrometric surveys of the Galaxy. For model fitting, the dark halo potential should be represented by distributions of the coefficients that can be used as priors in Bayesian inference from the data, rather than say a single number (the flattening) in a predetermined and unadaptable density law.
 
Of course, the shape of a dark halo depends on the nature of the dark
matter particle and on the extent of feedback processes~\citep[see
  e.g.,][]{Sellwood2004, Maccio2012}. The shape also depends on a host
of other factors, including the mass of the halo, its environment
(isolated versus group), its recent history (e.g., late infall of a
large subhalo) and the presence or absence of a disk. It will be
interesting to test our new basis function expansion method on the
full variety of numerically-constructed haloes, and understand how the
distributions of coefficients changes with the underlying physics. The
main impediment to efficient exploitation of these ideas is that so
few biorthogonal pairs are known. Our new discovery helps in this
regard, but it has not exhausted the supply of such expansions. We
will show in a later paper how to extend the methods in this paper to
other cosmologically-inspired dark-halo density laws.

\section*{Acknowledgements}
JLS and EL acknowledge the support of the STFC. We thank members of
the Institute of Astronomy Streams group for discussions and comments
as this work was in progress. \rd{The anonymous referee is thanked for
  a number of helpful suggestions.}



\bibliographystyle{mnras}
\bibliography{sources}



\bsp	
\label{lastpage}
\end{document}